\documentclass[%
 aip,
 amsmath,amssymb,
 reprint,%
]{revtex4-1}
\usepackage{xcolor}
\usepackage{graphicx}
\usepackage{dcolumn}
\usepackage{bm}

\usepackage[utf8]{inputenc}
\usepackage[T1]{fontenc}
\usepackage{mathptmx}
\usepackage{etoolbox}
\usepackage{physics}
\makeatletter
\def\@email#1#2{%
 \endgroup
 \patchcmd{\titleblock@produce}
  {\frontmatter@RRAPformat}
  {\frontmatter@RRAPformat{\produce@RRAP{*#1\href{mailto:#2}{#2}}}\frontmatter@RRAPformat}
  {}{}
}%
\makeatother
\usepackage{lipsum} 
\begin{document}
\preprint{AIP/123-QED}
	\title{Third harmonic-mediated amplification in TWPA}

	\author{E.~Rizvanov}
	\affiliation{Department of Experimental
		Physics, Comenius University, SK-84248 Bratislava, Slovakia}
	\email{emil.rizvanov@fmph.uniba.sk}
	\author{S.~Kern}
	\affiliation{Department of Experimental
		Physics, Comenius University, SK-84248 Bratislava, Slovakia}
	
	\author{P.~Neilinger}
	\affiliation{Department of Experimental
		Physics, Comenius University, SK-84248 Bratislava, Slovakia}

	\author{M.~Grajcar}
	\affiliation{Department of Experimental
		Physics, Comenius University, SK-84248 Bratislava, Slovakia}

\begin{abstract}

In Josephson Traveling-Wave Parametric Amplifiers, higher-order harmonics of the pump tone and its side-
bands are commonly present and typically regarded as parasitic. Consequently, most design efforts have
focused on suppressing these harmonics. In spite of that, motivated by transient simulations, we extend the
coupled-mode theory and demonstrate that the third harmonic can enhance
amplifier performance, improving both gain and bandwidth. We show that the recently developed plasma
oscillation-based amplifier is particularly well-suited for exploiting this effect. Its dispersion relation enables
us to observe the phenomenon in transient numerical simulations using JoSIM and WRspice. These simulations reveal improvement of the amplifier’s performance, specifically doubling the bandwidth and increasing the gain.

\end{abstract}
\pacs{}
\maketitle

\label{ch:intro}

Josephson Traveling Wave Parametric Amplifiers (JTWPAs) are a leading technology for ultra-sensitive detection of microwave signals, offering near-quantum-limited noise performance while maintaining high dynamic range and broad bandwidth spanning several gigahertz. They are essential for cutting-edge applications such as quantum computing, where they enable high-fidelity qubit readout, \cite{Devoret, Pagano_book} and quantum sensing, where detecting extremely weak signals is critical. \cite{Macklin, Pagano2} Recent rapid advancements lead to high-gain and low-noise compact devices for integration with quantum measurements. \cite{Planat2019, Samolov, Ranadive_2022, Zorin2016, Zorin2017, Dixon, Gaydamachenko2022, Kissling2023, Nilsson2023, Peatain2021, OBrien2014, White2015, Kissling2023, Peng_2022, Peng_FM, Macklin, BiSQUID}

The current understanding of their working principle is based on processes well known from nonlinear optics, \cite{PhysRevA.39.2519, First_JTWPA, PhysRevB.87.144301} where the strong pump tone interacts with the weak signal as they propagate along a nonlinear transmission line composed of Josephson junctions.
As a result, an idler mode is generated, enabling the energy transfer from the pump to the signal in a broadband and linear manner. The efficiency of this process depends critically on the phase relations between the interacting waves, as additional nonlinear phase shifts can degrade the gain and bandwidth.

To address the phase mismatch, dispersion engineering techniques - such as introducing band gaps or resonant features - have been employed to modify the dispersion relation. Placing the pump frequency near these features mitigates phase mismatch and enables optimal gain and bandwidth, although it makes impedance matching a more complex task, \cite{kow2025traveling} often leading to unwanted reflections. \cite{Kern2022}

In addition, harmonic generation can significantly degrade amplifier performance.\cite{9134828} In particular, third-harmonic generation (THG) of the pump introduces an additional noise channel and may promote shock-front formation, making it especially detrimental in certain amplifier designs.\cite{Eom2012,Dixon, Dixon_thesis,levochkinaTHG, Nilsom_Photonic_crystall, RenbergNilsson2023}. Again, dispersion engineering is typically utilized to suppress such generation by placing unwanted harmonics within bandgaps or by preventing the process through introduced phase mismatch.

The second and third pump's harmonics can be effectively suppressed using this technique. However, applying the same approach to sidebands generation, such as the pump-signal sum frequency $f_p+f_s$, is difficult as the suppression must be broadband to maintain the amplifier's bandwidth. These sidebands are shown to be particularly detrimental to the gain. Surprisingly, including the third harmonic, especially when its generation is phase-matched, can mitigate the negative effects of these sidebands. Furthermore, this leads to a novel amplification mechanism that relies on the aforementioned sidebands.

In this work, we report the study of this novel amplification mechanism that is observed in time-domain numerical simulations of a recently proposed traveling-wave parametric amplifier based on plasma oscillations (PTWPA). \cite{PTWPA} The transient analysis simulations, commonly used to handle the nonlinear process of higher order in
TWPAs,\cite{PTWPA, Pagano_SHG, Pagano_TWPA2, Dixon, Peatain2021, Kissling2023, Gaydamachenko2022, Peng_2022, Peng_FM, malnou2024RPMTWPA, Ranadive_2022, TWPAI} were performed in JoSIM \cite{JoSIMSite,  Delport2019} and WRspice \cite{ Spice_cite,133816} software. Motivated by this finding, we extended the coupled-mode equations (CME), following approaches similar to those in Refs. \onlinecite{Dixon, Dixon_thesis}, by incorporating the effects of third-order nonlinearity responsible for THG. Within this framework, we identified efficient, phase-matched THG as the key mechanism driving the observed amplification enhancement. Although the possibility of the beneficial role of THG in amplification was briefly mentioned in Ref.~\onlinecite{Peng_2022}, its contribution was not quantitatively analyzed or explicitly demonstrated. 

The underlying physical mechanism is identified utilizing a simplified model, revealing a distinct amplification channel involving higher-order sidebands and the third harmonic of the pump. 
We then extended the CME model to tailor it to the PTWPA's dispersion and demonstrate qualitative agreement with the simulation results. We discuss how the PTWPA design is particularly well-suited for realizing this amplification mechanism, as supported by dispersion analysis and frequency-domain modeling based on the extended CME. 

\begin{figure}
	\centering
	\includegraphics[width=8.6cm]{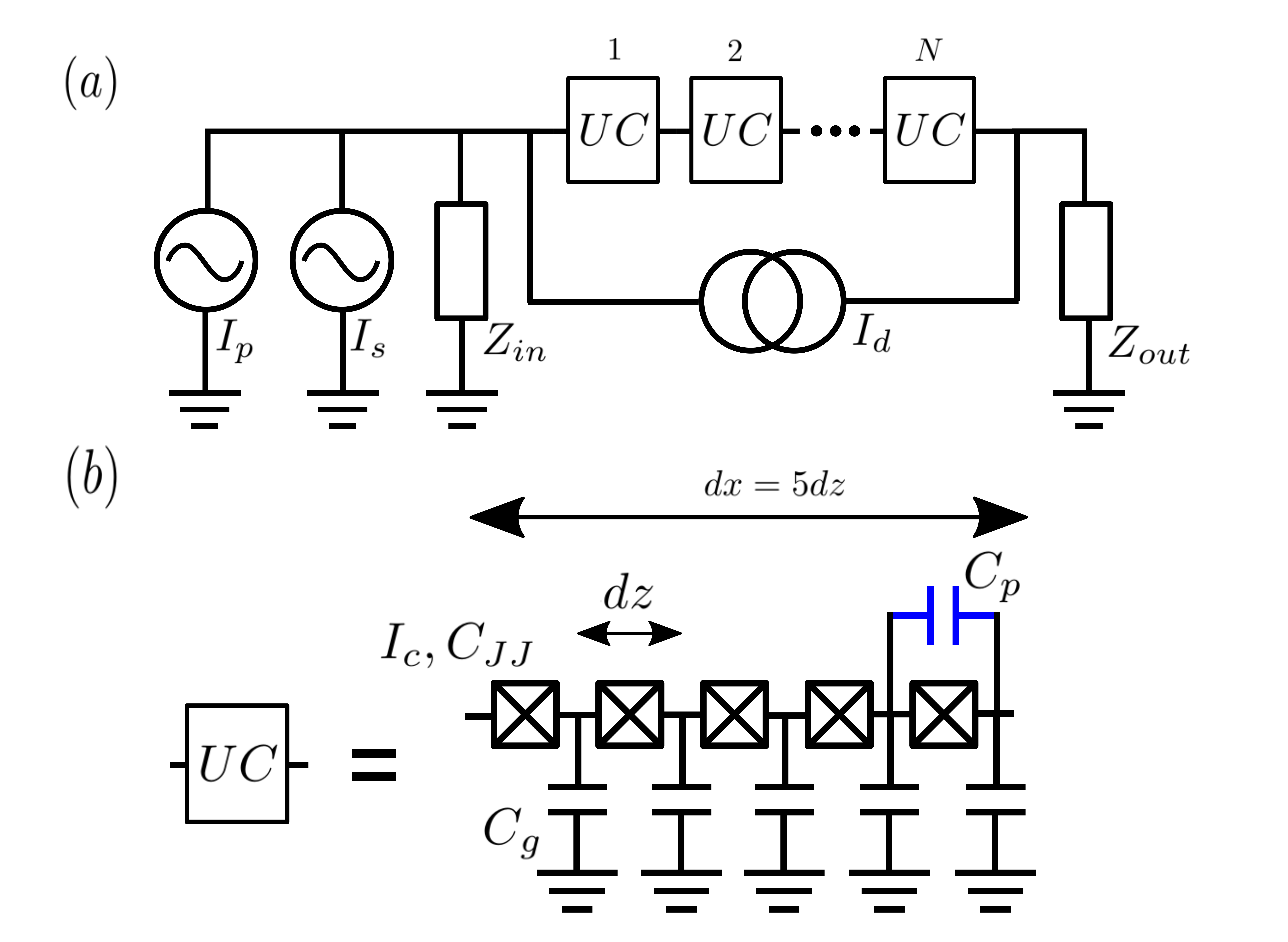} 
     \includegraphics[width=8.6cm]{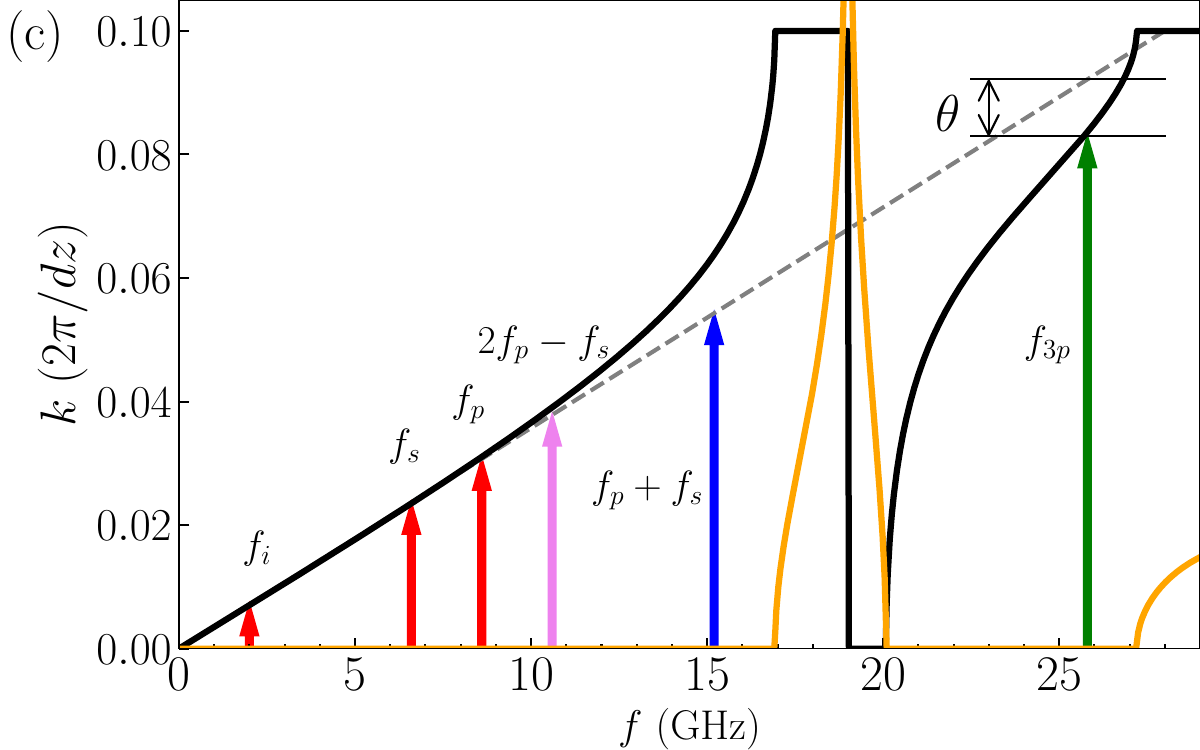}
    
	\caption{(a) Schematic  of PTWPA  implemented in
JoSIM and WRspice. It consists of 
an array of N identical unit cells (UC). $Z_{in/out} =\text{50 } \Omega$. (b) Schematic of PTWPA unit cell. Capacitor $C_{p} = \textrm{394 fF}$  (blue), responsible for plasma oscillation, is placed in parallel to every $5^{th}$ Josephson junction (black crosses) to modify dispersion. Parameters of RSCJ model of junctions are:$ \; I_{c} =2~\mu\textrm{A}, R = 550~\Omega, C_{JJ} =\textrm{12 fF}$. All ground capacitances have equal values: $ C_{g} =  \textrm{71.5 fF}$. In simulations, the whole circuit contains $N = 240$ unit cells. In CME calculations, we considered $dz = 5~\mu$\textrm{m}. (c) Dispersion curves for PTWPA unit cell; real (black) and imaginary part (orange) of $k$, respectively, obtained by ABCD matrix method. The wave vector is normalized to $2\pi/dz$, where $dz$ is the distance between adjacent junctions. DC current was set to $I_{d} = 0.8~\mu\textrm{A}$. Linear dispersion approximation utilized in CME to study the amplification process is shown as a grey dashed line. For third harmonic (green arrow) an additional phase shift $\theta$ is introduced for phase matching. Sidebands are indicated by purple and blue arrows.}
  \label{fig:JJ_RPM_pic}
\end{figure}

\begin{figure}
	\centering
    
    \includegraphics[width=8.6cm]{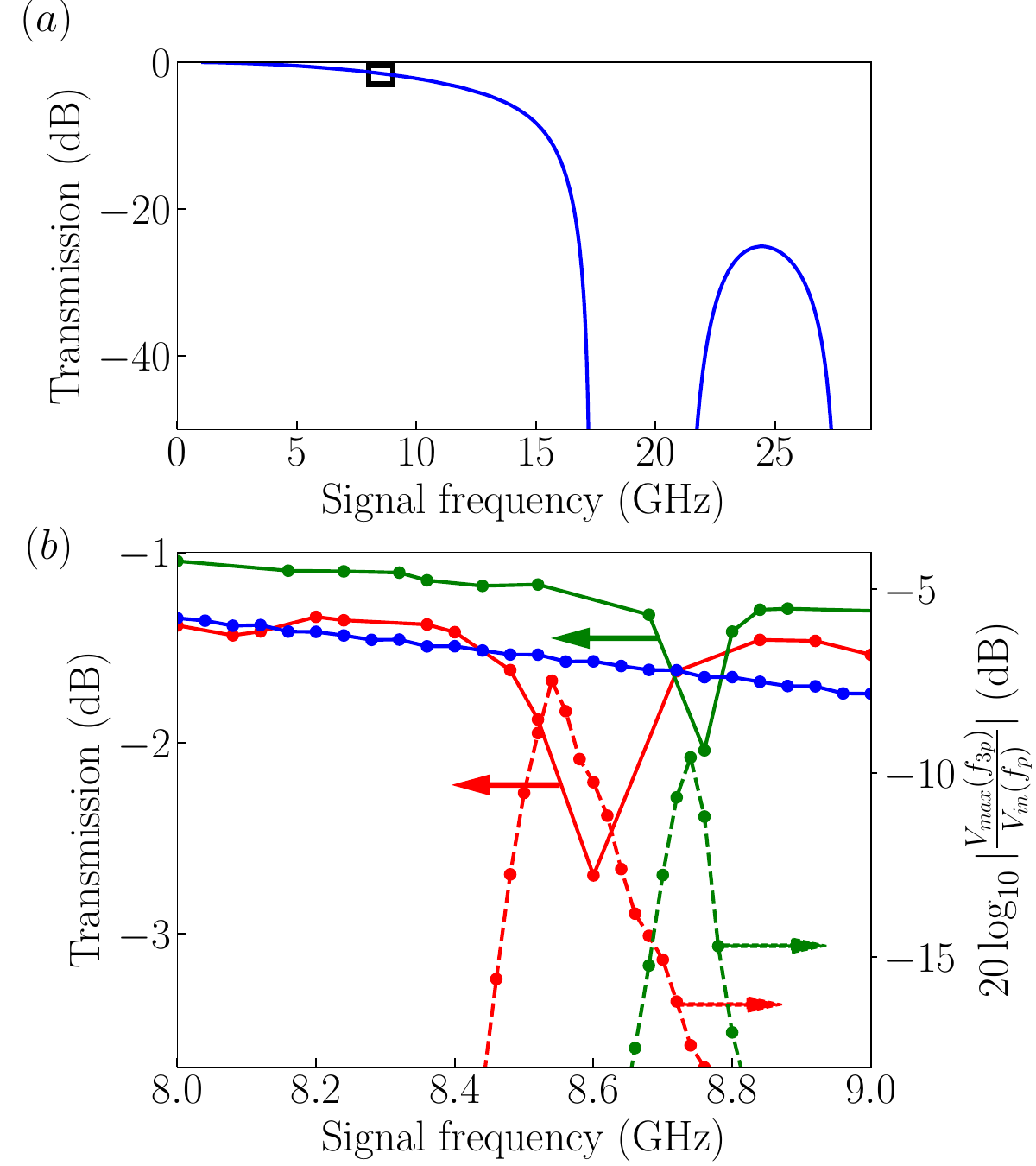} 
    
	\caption{ (a) PTWPA transmission spectra ($S21$) for single pump tone simulated in JoSIM as a function of signal frequency. (b) Strong signal transmission (green and red lines) and THG conversion gain (dashed lines). Parameters of the simulation are: $I_{p} = 1.6~\mu\textrm{A}$  (red line), $I_{p} = 1~\mu\textrm{A}$ (green line), $I_{p} = 0.01~\mu\textrm{A}$ (blue line), and $I_{d} = 0.8~\mu\textrm{A}$.}
  \label{fig:sweet_spot}
\end{figure}

\section{Time-domain simulation of gain in TWPA with plasma oscillation phase matching}

Recently, PTWPA was proposed, showing excellent performance in terms of gain, bandwidth, and reflections \cite{PTWPA} in transient simulations. 
The PTWPA utilizes dispersion influenced by Josephson plasma resonance, combined with a composite unit cell consisting of regularly alternating Josephson capacitances (see Fig.~\ref{fig:JJ_RPM_pic} (b)), resulting in a similar dispersion to one studied in Ref. \onlinecite{Tholen}. The 
dispersion relation, calculated using the ABCD matrix formalism, is shown in Fig.~\ref{fig:JJ_RPM_pic} (c) and the detailed derivation is provided in Appendix \ref{app:Transfer_matrix}. The dispersion exhibits a broad region of linear behavior governed by the Josephson inductance and the ground capacitance (indicated by the grey dashed line in Fig.~\ref{fig:JJ_RPM_pic} (c)). Near the plasma frequency, the dispersion bends and a bandgap opens, preventing wave propagation within the bandgap. Above this gap, wave propagation resumes, characterized by a band with a rapid increase of wavevector as a function of frequency.
In Ref.~\onlinecite{PTWPA}, the performance of PTWPA was numerically studied in the three-wave mixing  (3WM) regime, and the optimal pump frequency was found to be $f_{p} = 8.64 ~\textrm{GHz}$. 
Interestingly, closer examination of the transient simulation, presented in this work, revealed the system exhibits high gain in a relatively narrow band (tens of MHz) of pump frequency located in the linear segment of the dispersion at $\approx8.5~$GHz. 

This frequency band, which we will refer to as the "sweet spot", was found to be a region of highly efficient THG, as evidenced from the dip in the transmission spectrum of the strong pump in Fig.~\ref{fig:sweet_spot} and transmission spectra calculation presented in the Section~\ref{app:THG_generation}. Phase matching for the THG process is achieved because the third harmonic falls within the dispersion segment with the steep $k(\omega)$ slope, while the pump remains in the more linear region. Here, the THG phase matching is inferred from the steep increase of the third-harmonic amplitude without the rapid oscillations (black solid line in Fig. 3 (a)). Additionally, under these conditions, the second harmonic lies within the bandgap, thereby not interfering with the observed effect.

\begin{figure}
	\includegraphics[width=8.6cm]{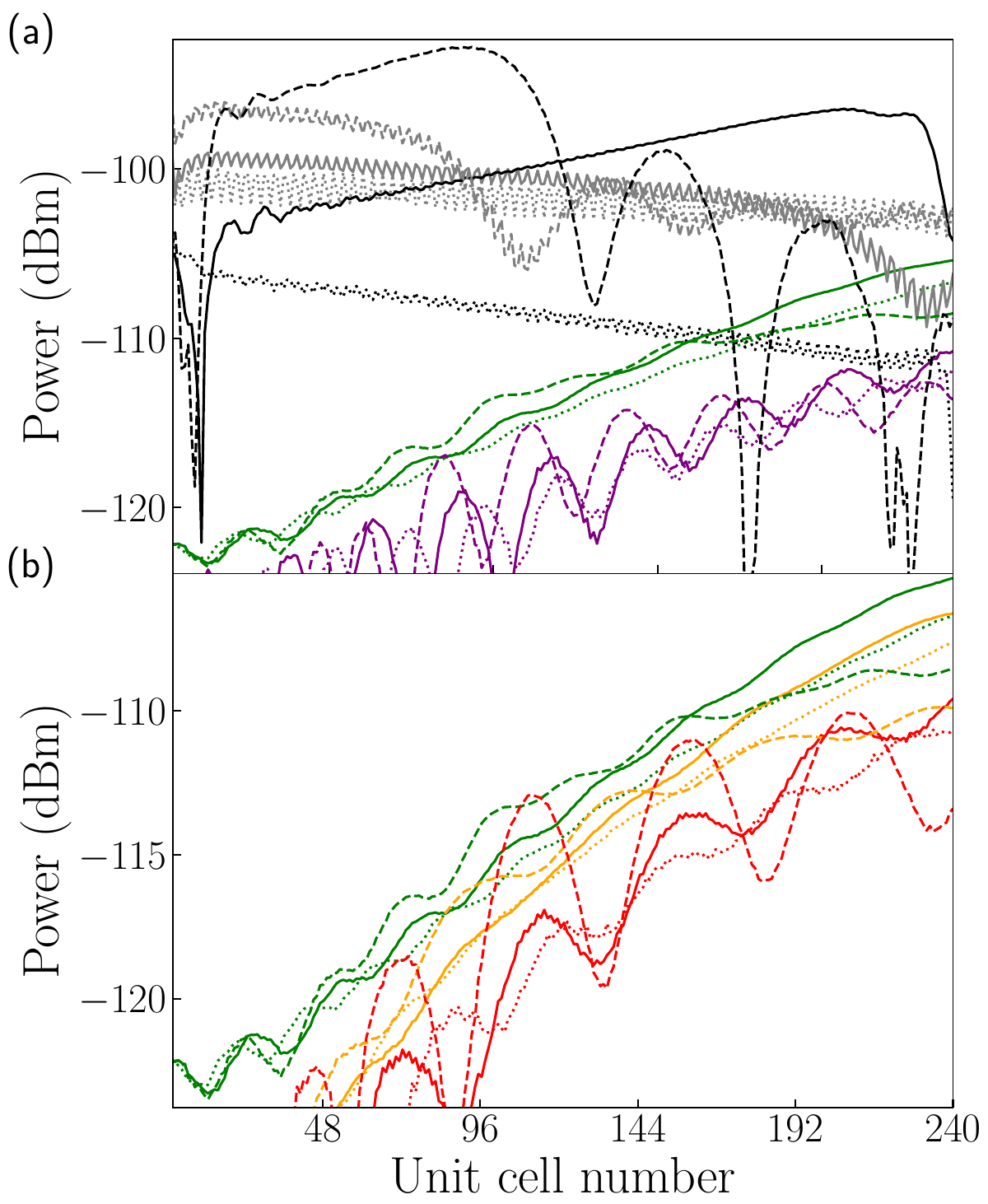}
	\caption{Power flow of propagating plane waves for relevant modes under three pump configurations simulated in JoSIM, each corresponding to a different third-harmonic generation efficiency: $f_{p} =  8.5 \textrm{ GHz}, I_{p} = 1.8~\mu\textrm{A}$  (dashed line); $f_{p} =  8.64 \textrm{ GHz}, I_{p} = 1.6~\mu\textrm{A}$  (solid line); $f_{p} = 8.8 \textrm{ GHz}, I_{p} = 1.7~\mu\textrm{A}$ (dotted line). Respective modes are a) $f_{3p}$ (black lines), $f_{2p}$ (gray lines), $f_{s}$ (green lines), $f_{p+s}$ (purple lines); b) $f_{s}$ (green lines), $f_{2p-s}$ (red  lines), $f_{p-s}$ (orange lines). The remaining parameters of the simulation are: $f_{s} = 4.8 \textrm{ GHz}; I_{s} = 0.01~\mu\textrm{A}, I_{d} = 0.8~\mu\textrm{A}$.} 
    \label{fig:Power_propagation_JoSIM}
\end{figure}

\begin{figure}[h]

    \includegraphics[width=8.6cm]{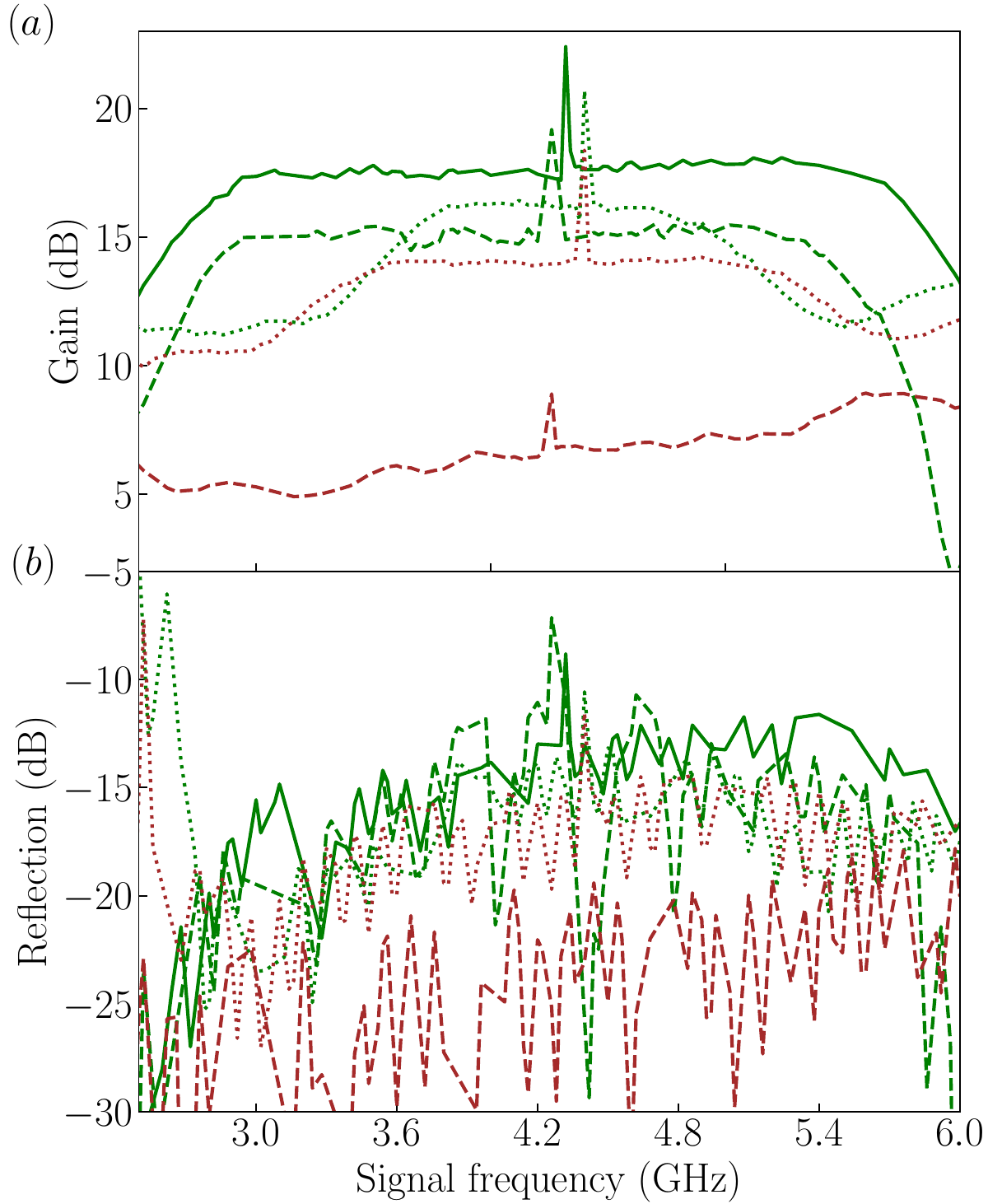}
    \caption{\label{fig:Gain_fp_comparing}  Gain (a) and reflections (b) simulated in JoSIM for three pump configuration cases presented in Fig.~\ref{fig:Power_propagation_JoSIM}. Parameters of the simulation are: $f_{p} = 8.5 \textrm{ GHz}, I_{p} = 1.8~\mu\textrm{A}$ (green dashed line), $I_{p} = 1.6~\mu\textrm{A}$ (brown dashed line); $f_{p} =  8.64 \textrm{ GHz}, I_{p} = 1.6~\mu\textrm{A}$ (green solid line); $f_{p} = 8.8 \textrm{ GHz}, I_{p} = 1.7~\mu\textrm{A}$ (green dotted line), $I_{p} = 1.6 ~\mu\textrm{A}$ (brown dotted line);  $I_{s} = 0.01~\mu\textrm{A}, I_{d} = 0.8~\mu\textrm{A}$.
} 
\end{figure}

In Ref.~\onlinecite{PTWPA}, the optimal pump frequency, in terms of maximal gain and low reflections, was found to be $f_{p} = 8.64~\textrm{GHz}$, which, indeed falls in the sweet spot. We note that here the length of the PTWPA was chosen to be $N = 240$ unit cells instead of 400. \cite{PTWPA}
The role of the sweet spot could be unnoticed, as the basic 3WM theory can provide a similar gain profile to the simulated ones. However, an  investigation, presented in this work, reveals that the presence of the sweet spot is a result of a more complex amplification mechanism driven by the third harmonic of the pump, $f_{3p}$. Throughout the paper, we adopt the notation $f_{mp} = mf_p$ and $f_{mp \pm s} = mf_p \pm f_s$, where $m$ denotes the $m$-th harmonic of the pump.

The importance of the third-harmonic generation (THG) is evident from the individual harmonic propagation simulated in JoSIM for the pump near the sweet spot, as shown in Fig.~\ref{fig:Power_propagation_JoSIM}. 
Sweeping the pump frequency around the sweet spot reveals strong dependence of the third harmonic generation efficiency (see black lines in Fig.~\ref{fig:Power_propagation_JoSIM}, solid: efficient THG, dashed: poor THG efficiency). Furthermore, it was observed that the signal gain increase (see green lines in Fig.~\ref{fig:Power_propagation_JoSIM}) is correlated with the enhancement of the third harmonic, which enables shorter PTWPA design. 
Even more striking observation is the considerable enhancement of the amplifier's bandwidth as shown in Fig.~\ref{fig:Gain_fp_comparing}. For strongly suppressed THG (e.g., at $f_{p} = 8.8\textrm{ GHz}$, where $f_{3p}$ is in the cutoff region, black dotted line in Fig.~\ref{fig:Power_propagation_JoSIM} (a)), the amplifier bandwidth is approximately half that for the pump in the sweet spot, where third harmonic of the pump is efficiently generated.

Notably, there is a clear correlation between the phase matching of the third harmonic (e.g., the third harmonic amplitude increases exponentially without any oscillations) and the amplifier gain: the better the phase matching, the stronger the THG and the higher the resulting gain. Vice versa, when THG becomes suppressed after 100 unit cells, the gain is suppressed as well (dashed lines in Figs.~\ref{fig:Power_propagation_JoSIM} and \ref{fig:Gain_fp_comparing}). This observation indicates that, although THG is typically detrimental to amplification, \cite{levochkinaTHG, Nilsom_Photonic_crystall, RenbergNilsson2023} it can also contribute positively by enhancing key amplifier parameters such as gain and bandwidth.

Simulation results obtained from both JoSIM and WRspice indicate that while the third harmonic is efficiently generated, the second harmonic is significantly attenuated - even though both fall above the plasma frequency $f_{plasma} \approx 19~\textrm{GHz}$. This attenuation is attributed to the periodic modulation of the parallel capacitance, which gives rise to the stopband above the cutoff frequency (see Fig.~\ref{fig:JJ_RPM_pic}(c)). Transmission above the plasma frequency of a Josephson Junction array with a similar dispersion relation was measured in Ref.~\onlinecite{Tholen}.

Finally, the transient simulation does not show front formation, a potentially critical effect in nonlinear systems with higher harmonics. A potential increase of the noise temperature, even if not directly verified, should not be significant, as the third harmonic is suppressed (-20 dB (see Fig. \ref{fig:sweet_spot} (a))). Moreover, the amplifier gain is only marginally affected when the spread in critical current remains below 7.5 \% (Appendix \ref{ch:Parameter_variation}).

\section{Simplified CME analysis of parametric amplification via third harmonic generation}
\begin{figure}[!ht]
	\centering
	\includegraphics[width=8.6cm]{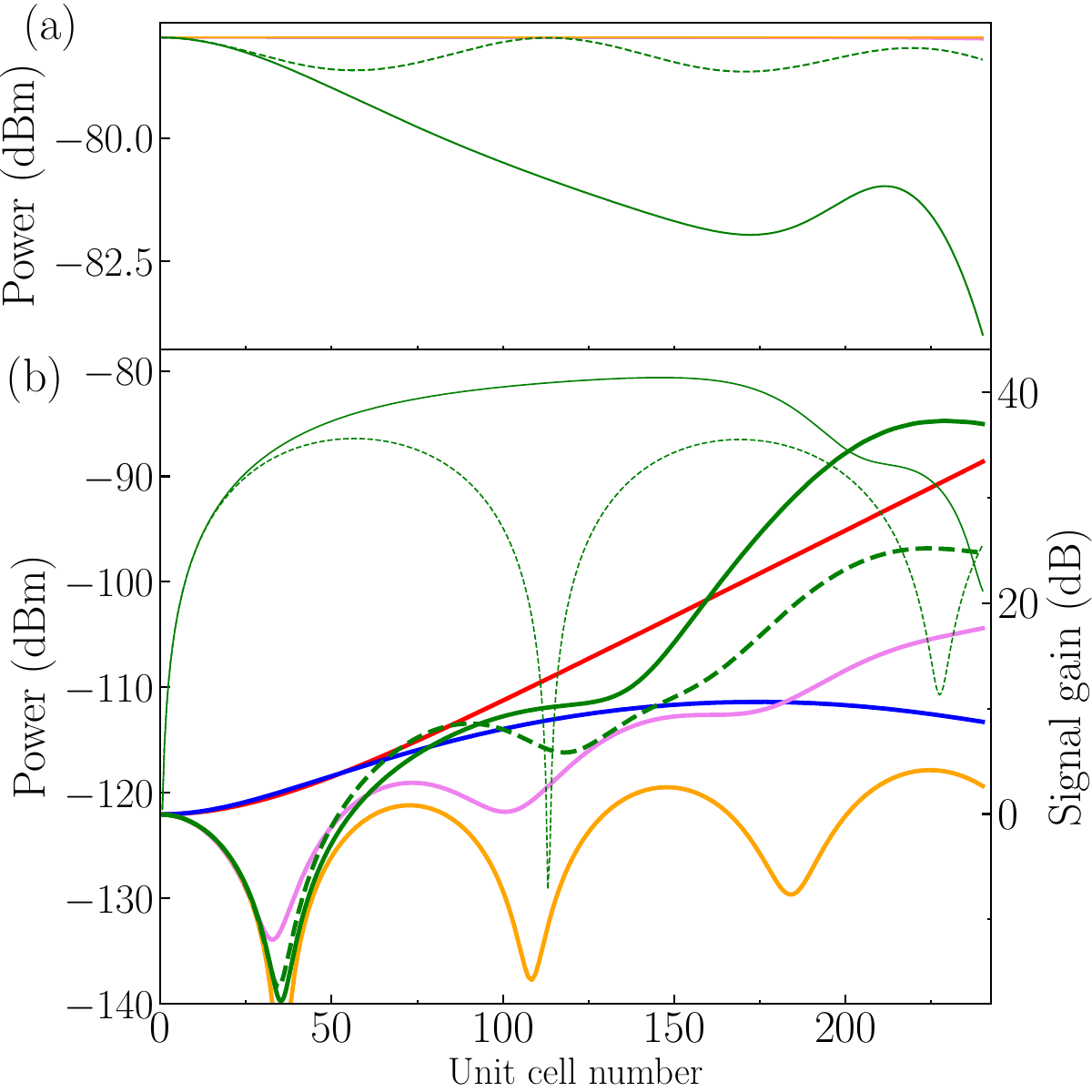}
    \includegraphics[width=8.6cm]{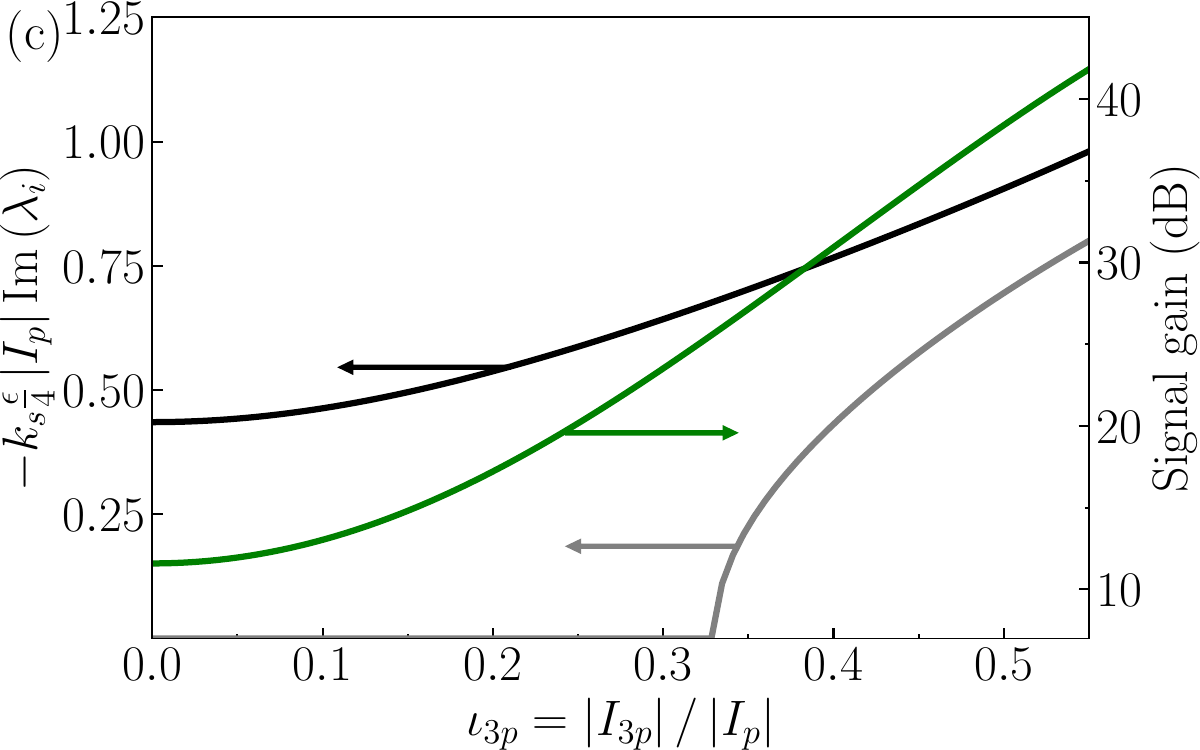}
	\caption{(a) Node evolution of the pump tone calculated from the basic CME including: pump, signal, and idler (red); addition of a non–phase-matched third harmonic (green dashed); and addition of a phase-matched third harmonic (green solid). (b) Thick solid lines represent signal gain according to CME with various modes: 3WM theory with signal, idler, pump (red), with additional 4WM idler $2f_p-f_s$ (purple), or with additional $f_p+f_s$ tone (blue), and with both sidebands (orange). Process with both sidebands and third harmonic  not phase matched (green dashed line), and phase matched third harmonic  (green solid line). Thin green lines are evolutions of the respective $f_{3p}$ mode. (c) Dependence of the negative normalized imaginary parts of the eigenvalues (black and grey curves) of matrix describing the simplified CME (left axis) on the third harmonic amplitude and corresponding signal gain (green line) - right axis. The calculations used parameters: $f_p=9~$GHz, $f_s=4.8~$GHz, and $I_p=I_d=0.4I_c$, which for $I_{c} = 2 ~\mu\textrm{A}$ corresponds to $I_p=I_d = 0.8 ~\mu\textrm{A}$, same as in transient simulations. }
  \label{fig:3wm_thg}
\end{figure}

For a qualitative understanding of the nonlinear processes in the studied TWPA, the CME is utilized. In CME, the junction array is approximated as a continuous nonlinear waveguide, which can be described by the nonlinear wave equation governing the current $I(x,t)$ as a function of space and time:\cite{malnou2024RPMTWPA,Malnou_TWPA_CME,Kern2022}
\begin{equation}
\label{eq:wave_eq}
 v_{p}^{2} \frac{\partial^{2}I(x,t)}{\partial x^{2}} - \frac{\partial^{2}I(x,t)}{\partial t^{2}} = \frac{\partial^{2}}{\partial t^{2}} \left(\frac{1}{2} \epsilon I(x,t)^{2} +\frac{1}{3}\xi I(x,t)^{3}\right),
\end{equation}
where the coefficients of second and third-order nonlinearity are given by:
\begin{equation}
    \epsilon = \frac{2I_{d} }{(I_{*}^{2} + I_{d}^{2})},\  \xi =\frac{1}{(I_{*}^{2} + I_{d}^{2})},\   
     I_{*} = \sqrt{2}I_{c}.
\end{equation}
Here, $I_c$ is the junctions critical current and $I_d$ is the applied DC current which enhances the second-order nonlinearity while suppressing the third-order contribution at the RHS of wave equation (\ref{eq:wave_eq}). 

The equation is solved using a plane wave ansatz with spatially dependent current amplitudes:
\begin{equation}
\label{eq:wave_ansatz}
    I(x,t) = \sum_\alpha \frac{1}{2}I_{\alpha}(x)\exp(i(k_{\alpha}x - \omega_{\alpha} t))  + C.C.,
\end{equation}
for a certain set of modes $\alpha$, which define the order of the CME. CME of $n$-th order (CME-$n$)  contains all pump-mediated mixing tones up to and including the $n$-th pump's harmonic.\cite{Dixon}

CME is then based on the evolution of the amplitudes $I_\alpha(x)$ resulting from nonlinear effects. The nature of the plane waves $\alpha$ is encoded in their dispersion $k(\omega)$, which enters the theory as phase mismatches for each process, represented by mixing terms in ODE system resulting from substituting the ansatz in the wave equation. Finally, the power of each harmonic can be calculated as
\begin{equation}
P_{\alpha} = 10 \log_{10} \left( \frac{\frac{1}{2} \left| I_{\alpha} \right|^2 Z}{1~\textrm{mW}} \right),
\end{equation}
where $Z = 50~\Omega$ is the characteristic impedance.

It turns out that the influence of the third harmonic can be captured within this framework, provided CME is extended beyond the standard modes, namely: pump $f_p$, signal $f_s$, and idler $f_i$, to also include the sidebands $f_{p+s}$ and $f_{2p-s}$, and the third harmonic $f_{3p}$.
Before presenting a detailed model of the simulated structure and its dispersion via the CME-3 (i.e., for the set of six modes $\{f_p, f_s, f_i, f_{p+s}, f_{2p-s}, f_{2p},f_{2p+s}, f_{3p-s}, f_{3p}\}$) analyzed in Section \ref{app:THG_generation} and Appendix \ref{app:CME3}), a simplified system is analyzed in order to identify and understand the individual effects responsible for the gain. This approach requires various approximations and is therefore purely conceptual, without any quantitative conclusions. However, it sheds light on the mechanism behind the observed effects of THG.

In the standard three-wave mixing (3WM) model involving only the pump, signal, and idler (i.e., CME-1), the signal power exhibits decent growth along the TWPA, even without any phase matching. This is an advantage of 3WM systems, which suppresses the detrimental effects of nonlinear phase modulation. \cite{Malnou_TWPA_CME} Extending the model, however,  to include the additional sidebands results in a substantial gain decrease  (blue and violet curves in Fig. \ref{fig:3wm_thg} (b)),\cite{Dixon,Dixon_thesis} even down to zero
when both $f_p+f_s$ and $2f_p-f_s$ are present (yellow curve in Fig. \ref{fig:3wm_thg} (b)).

THG, which occurs naturally in systems governed by the third order nonlinearity, by itself depletes the pump power,\cite{OBrien2014} thus reduces the gain
and is typically intentionally suppressed. Nevertheless, the strength of the depletion is highly dependent on the underlying dispersion relation (see pump depletion in the cases without, with non-phase-matched, and phase-matched third harmonic in Fig. \ref{fig:3wm_thg} (a)). By adopting a linear dispersion relation with an additional phase offset $\theta$ applied to the third harmonic, we can track how the phase matching affects its generation efficiency and finally the signal gain.

Surprisingly, including the third harmonic in the coupled-mode model leads to a recovery of signal gain (green dashed line in Fig.~\ref{fig:3wm_thg}(b), even in the presence of sidebands and without THG phase matching. However, when the phase offset $\theta$ is tuned in order to improve its phase-matching, the third harmonic generation is enhanced (thin green line), resulting in a noticeable improvement in signal amplification (thick solid green line). This feature reproduces the results observed in the time-domain simulation presented in the previous section.

The distribution of pump power among the modes is analyzed withis the described CME models. In the simplest CME-1 model (pump, signal, idler), the pump power changes by less than 1 \%. Including the third harmonic increases pump depletion to approximately 14 \%, with 13 \% transferred to the third harmonic and the remainder distributed among the signal, idler, and sidebands. For a phase-matched third harmonic, the pump depletion is even more efficient, reaching 75 \%, of which 54 \% goes to the third harmonic, while the rest is approximately equally shared among the other modes.

To track the mechanism behind the observed phenomenon, we employed CME-3 and systematically removed individual interaction terms on the right-hand side of (see Appendix \ref{app:CME3}) and examined how these changes affected the correlation between the signal gain and THG.  The terms whose exclusion eliminated the correlation between third-harmonic generation (THG) and gain constitute the minimal set of equations governing the amplification dynamics, as detailed in Eqs.~(\ref{eq:dp3p}-\ref{eq:ddpms}) of Appendix C. The essential modes included in this reduced model are the signal, the idler, the pump, the third harmonic, and the two sidebands at frequencies $f_p+f_s$ and $2f_p-f_s$.

Further insight can be gained by applying the common undepleted-pump approximation $d\left|I_p\right|/dz=0$, and extending it to the third harmonic, $d\left|I_{3p}\right|/dz=0$. Under these assumptions, only the phases of these modes evolve due to self- and cross-phase modulation. These modulations can be captured analytically (see Eqs.~(\ref{eq:p_3p_sub}-\ref{eq:Y2ps}) in the Appendix \ref{app:dominant}) and substituted back into the remaining coupled mode equations, with the amplitude of the pump and the third harmonic as parameters, introducing a new set of amplitudes $Y_s, Y_i, Y_{p+s}, Y_{2p-s}$.

The resulting model takes the form
\begin{equation}
\frac{\partial}{\partial x}
\begin{pmatrix}
	Y_s\\
    Y_i\\
	Y_{p+s}\\
	Y_{2p-s}
\end{pmatrix} = i\frac{k_s\epsilon \left|I_p\right| }{4}\hat{A}
\begin{pmatrix}
    Y_s\\
    Y_i\\
	Y_{p+s}\\
	Y_{2p-s}
\end{pmatrix},
\end{equation}
where the interaction matrix is given as
\begin{equation}
\hat{A} = 
\begin{pmatrix}
\frac{k_p}{k_s}\iota_{p}& 1&1&0\\
-\frac{k_{p-s}}{k_s}&-\frac{k_p}{k_s}\iota_{p}&                0 &0\\
\frac{k_{p+s}}{k_s}&                0& 3\frac{k_p}{k_s}\iota_{p}& \frac{k_{p+s}}{k_s} \iota_{3p}\\
0&0&-\frac{k_{2p-s}}{k_s}\iota_{3p}&\delta
\end{pmatrix},
\end{equation}
with $\delta = 3\frac{k_p}{k_s}\iota_p\left(1-2\iota_{3p}^2\right) - \frac{\xi\theta}{k_s\epsilon^2\iota_p}$, $\theta$ is the dispersion offset of the third harmonic, $\iota_{p}=\frac{\xi}{4\epsilon}I_p$, and $\iota_{3p}=I_{3p}/I_p$.
Diagonalizing the matrix $A$ provides the solution in the form
\begin{equation}
\vec{Y}(x) = \sum_i c_i \vec{v}_i e^{i\frac{k_s\epsilon}{4}I_p \lambda_i x},
\end{equation}
where $\vec{v}_i$ and $\lambda_i$ are the eigenvectors and the eigenvalues of $\hat{A}$, respectively, and the coefficients $c_i$ satisfy the initial condition. The real part of $\lambda_i$ corresponds to oscillatory behavior, whereas the imaginary part corresponds to amplification ($\Im(\lambda_i)<0$) or attenuation ($\Im(\lambda_i)>0$).

As shown in Fig.~\ref{fig:3wm_thg}(c), the gain increases significantly with the third-harmonic amplitude $\iota_{3p}$ (green curve, left axis). Simultaneously, one of the eigenvalues develops an increasingly negative imaginary part (black curve, right axis), which strongly correlates with the signal gain.

Analysis of the interaction matrix reveals that the third harmonic plays a crucial role in mediating coupling to the $f_{p+s}$ and $f_{2p-s}$ sidebands. Specifically, it introduces an additional pathway for generating these modes. When the third harmonic is weak, generation of $f_{p+s}$ occurs primarily at the expense of the signal, thereby reducing the overall gain. This process is expressed by the elements $\hat{A}_{13}$, $\hat{A}_{13}$. In contrast, the $\hat{A}_{34}$ and $\hat{A}_{43}$ elements, both proportional to the third harmonic strength,
describe the generation of sidebands at the expense of the third harmonic. The generation of the sum mode 
$f_{p+s}$ in this way alters the relative phase of the involved modes and prevents signal depletion.
As a result a strong third harmonic facilitates simultaneous exponential growth of all involved modes, leading to enhanced amplification; the effect clearly observed in the transient simulations.

\section{Complete CME analysis}
\label{ch:CME}

The described effects remain relevant in the studied design, despite its increased complexity, as captured by the dispersion relation shown in Fig. \ref{fig:JJ_RPM_pic} (c). Namely, both the sidebands and the third harmonic exhibit significant dispersion, and second harmonic is present, although it is suppressed.

First, we demonstrate that the amplification sweet spot is an intrinsic feature of the studied PTWPA design. This behavior is well reproduced by the CME and is clearly linked to the third harmonic generation.

Second, we apply the extended CME in the three-wave mixing regime, using the dispersion relation given by the ABCD matrix formalism (see Appendix \ref{app:Transfer_matrix}) to calculate the signal gain. This analysis confirms the robustness of the identified amplification mechanism identified using the simplified dispersion model in the previous section.

\subsection{\label{app:THG_generation} Generation of the pump's third harmonic}

To explain the nature of the sweet spot, i.e., a narrow band for the pump frequency, where the gain was observed, and to support the assumption regarding the impact of the third harmonic on the transmission (see Fig.~\ref{fig:sweet_spot}), we first analyze the case involving only two waves; the pump and its third harmonic:
\begin{equation}
\begin{aligned}
    I &=  \frac{1}{2}I_{p}(x)\exp(i(k_{p}x - \omega_{p} t))  \\ &+\frac{1}{2}I_{3p}(x)\exp(i(k_{3p}x - 3\omega_{p} t))  + C.C.
 \end{aligned}
\end{equation}
Substituting this anzatz to Eq. 1, we obtain: 
\begin{equation}
\label{eq:SPM_1}
\begin{aligned}
\frac{dI_p}{dx} = i \frac{k_p \xi }{24}  \big(
    I_{p}(6   \left|I_{3p}\right|^2  +3  \left|I_p\right|^2) + 
    3  I_{3p}  e^{i x (k_{3p} - 3k_p)} (I_p^{*})^2
    \big),
\end{aligned}
\end{equation}
\begin{equation}
\label{eq:SPM_2}
\begin{aligned}
\frac{dI_{3p}}{dx} = i \frac{k_{3p}\xi }{24} \big(
     3  I_{3p}(\left|I_{3p}\right|^2 +
     \left|I_{p}\right|^2) +   I_p^3  e^{i x (-k_{3p} + 3k_p)} \big).
\end{aligned}
\end{equation}
\begin{figure}[!ht]
	\includegraphics[width=8.6cm]{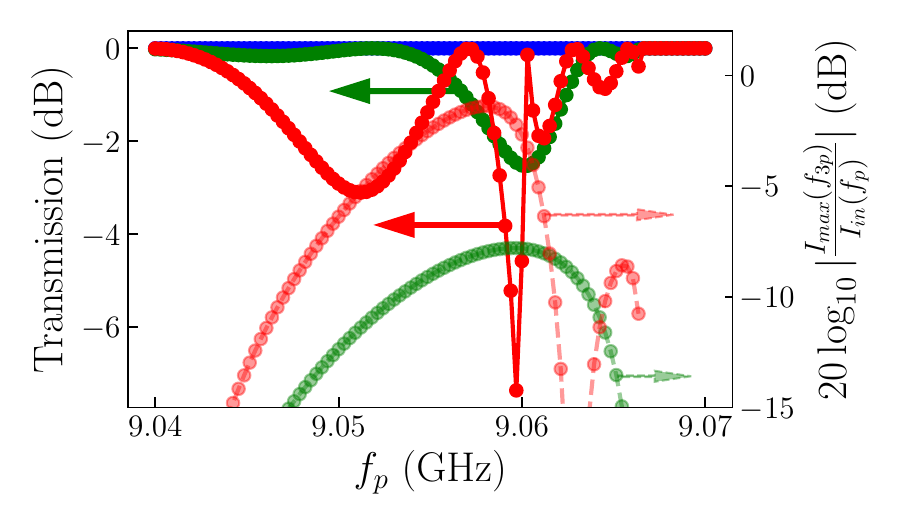}
	\caption{ PTWPA Transmission spectra ($S_{21}$ (left axis)) of single pump tone given by Eq.(\ref{eq:SPM_1}) and Eq.(\ref{eq:SPM_2}) for different powers. Parameters of the simulation are: $I_{p} = 1\textrm{ uA} ~(\text{red line}), ~I_{p} = 0.5\textrm{ uA} ~(\text{green line}), ~I_{p} = 0.01\textrm{ uA} ~(\text{blue line}), ~I_{d} = 0.8\textrm{ uA}$. The lighter shade of each color indicates the corresponding third-harmonic maxima (right axis).} 
	\label{fig:Sweet_spot_CMT}
\end{figure}

The dispersion (Fig.~\ref{fig:JJ_RPM_pic}(c)) allows us to set $f_{3p}$ to be in the bandgap (see Fig.\ref{fig:sweet_spot}), where $Im(k_{3p}) \neq 0$. We handled this by assuming $I_{3p} = 0$ in Eqs.~\ref{eq:SPM_1},\ref{eq:SPM_2}. Solving this system of equations, we obtain a similar sweet spot (Fig.\ref{fig:Sweet_spot_CMT}) as was obtained in  JoSIM simulations  (Fig.~\ref{fig:sweet_spot}). The sweet spot in this case is noticeably narrower than in simulations (Fig.~\ref{fig:sweet_spot}). This may be caused by the approximate dispersion relation (see Section \ref{app:Transfer_matrix}).

\subsection{Signal gain}
\label{s:kw}

\begin{figure}[!ht]
	\includegraphics[width=8.6cm]{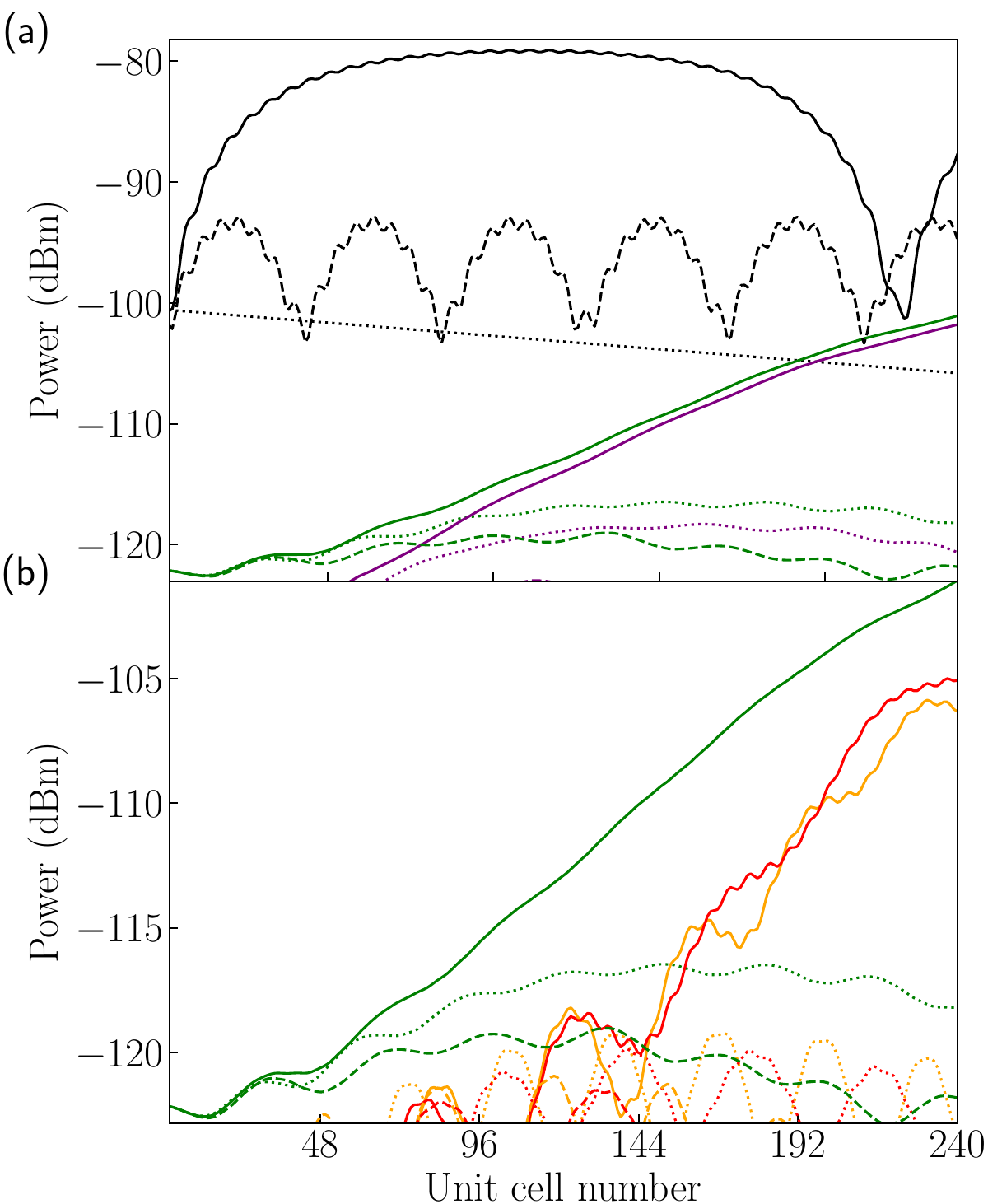}
	\caption{Power flow for relevant modes according to CME-3. a) $f_{3p}$ (black line), $f_{s}$ (green line), $f_{p+s}$ (purple line); b) $f_{s}$ (green line), $f_{2p-s}$ (red  line), $f_{p-s}$ (orange line). Calculation was done by  Explicit Runge-Kutta method of order 3(2). Parameters of the simulation are: $f_{s} = 4.8 \textrm{ GHz}; f_{p} =  8.912 \textrm{ GHz} ~(\text{dashed line}); f_{p} =  9.062 \textrm{ GHz} ~(\text{solid line}); f_{p} = 9.212 \textrm{ GHz} ~(\text{dotted line}),  I_{p0} = 1~\mu\textrm{A}, I_{s0} = 0.005 ~\mu\textrm{A}, I_{d} = 0.8~\mu\textrm{A}$. From Eq.\ref{eq:plas_dis}: $\Delta k_{3p}= 0.0008$ for $fp = 9.061 \textrm{ GHz}$; $\Delta k_{3p}= 0.0352$ for $fp = 8.911 \textrm{ GHz}$; $\Delta k_{3p}=0.0017$ for $fp =  9.211 \textrm{ GHz}$, where $\Delta k_{3p}= k_{3p} -3k_{p}$. Initial conditions for whole set of harmonics:
 $I_{s}(0) = I_{s0}, I_{p}(0) = I_{p0},  I_{2p}(0) = 0.08 I_{p}(0), I_{3p}(0) = 0.06 Ip(0); ~I_{\alpha}(0) = 0, \alpha \in \{p-s,p+s,2p-s,2p+s,3p-s\}$.}
	\label{fig:Power_propagation_CME}
\end{figure}

Motivated by our numerical simulations we included the second $f_{2p}$ and the third harmonic $f_{3p}$ of the pump, as well as the pump-mediated frequency tones:  $f_{p-s}$, $f_{p+s}$, $f_{2p-s}$, $f_{2p+s}$, $f_{3p-s}$. We assumed the signal to be much weaker than the pump, neglecting the second and the higher-order sidebands as in Ref.~\onlinecite{Side_bands}:
\begin{equation}
\label{eq:wave_ansatz}
\begin{aligned}
    I = \sum_{\alpha } \frac{1}{2} I_{\alpha}(x)\exp(i(k_{\alpha}x - \omega_{\alpha} t)) + C.C, \\
    \alpha \in \{s,p-s,p,p+s,2p-s,2p,2p+s,3p-s,3p\}.
\end{aligned}
\end{equation}

Fig.~\ref{fig:Power_propagation_CME}(a) shows that the qualitative behavior predicted by the extended CME is consistent with the simulation results. In particular, smaller values of $\Delta k_{3p}$ lead to higher gain and a significant increase in the third harmonic amplitude. This is strongly correlated with the enhanced signal gain as illustrated by the comparison of solid, dashed, and dotted green lines in Fig.~\ref{fig:Power_propagation_CME}(a). Moreover, the result for sideband $f_{p+s}$, which increases along the signal in the case of the matched third harmonic (see Fig.~\ref{fig:Power_propagation_CME}(a)), supports the analysis presented in the previous section. 

The discrepancy between the theory and the simulations may be caused by a restricted number of harmonics considered in the theory \cite{Dixon,Dixon_thesis} as well as by neglecting the reflections \cite{Kern2022,malnou2024RPMTWPA}, and the power dependence of the dispersion itself \cite{Planat2019}.

\label{ch:sim}
\section{Conclusion}
\label{ch:con}

In this study, we present numerical evidence that proper phase matching of third-harmonic generation (THG) can substantially enhance the performance of traveling-wave parametric amplifiers (TWPAs). Specifically, phase-matched THG results in a twofold increase in bandwidth and a significant gain enhancement. Conversely, we find that amplification is more efficient in the absence of THG than when THG is present but not phase-matched. These findings are consistent with prior studies and offer a deeper understanding of the influence of THG on gain and bandwidth characteristics.

To support these results, we extend the coupled-mode theory to incorporate both second- and third-order nonlinearities, analyzing interactions up to the generation of the third pump harmonic. Within this framework, we identify the generation of the \( f_{p+s} \)  sideband as the most detrimental process to amplification. We demonstrate that phase-matched third-harmonic generation can effectively mitigate this suppression, thereby improving the overall gain profile of the TWPA.

\subsection{Acknowledgments}
This work was supported in part by the Slovak Research and Development Agency under contract No. APVV-20-0425, the SPS Programme NATO grant number G5796, and the Comenius University in Bratislava through the CLARA@UNIBA.SK high-performance computing facilities, services, and staff expertise at the Centre for Information Technology (https://uniba.sk/en/HPC-Clara). The authors thankfully acknowledge the computer resources at MareNostrum5 and the technical support provided by Barcelona Supercomputing Center (RES-FI-2025-1-0044). Part of the research results was obtained using the computational resources procured in the national project National competence centre for high performance computing (project code: 311070AKF2) funded by European Regional Development Fund, EU Structural Funds Informatization of society, Operational Program Integrated Infrastructure.

The authors would like to thank Roman Martoňák, Matej Badin and David Vrba for their assistance in setting up the simulation platform and for sharing access to their numerical cluster.


\clearpage

\onecolumngrid

\appendix

\section{\label{app:Transfer_matrix}  Dispersion relation} 

The ABCD matrix of a waveguide element containing a single Josephson junction in a unit cell in Fig.~\ref{fig:JJ_RPM_pic}(b) is:
\begin{equation}
  M_{jj} = \begin{pmatrix}
        1 - \frac{\omega^{2}LC_{g}}{1 -\omega^{2}LC_{JJ}} & \frac{i\omega L}{1 -\omega^{2}LC_{JJ}}\\
       i \omega C_{g} & 1 
    \end{pmatrix},
\end{equation}
while for the junction with additional capacitance, the ABCD matrix reads:
\begin{equation}
  M_{p}  =  \begin{pmatrix}
        1 - \frac{\omega^{2}LC_{g}}{1 -\omega^{2}L(C_{JJ}+C_{p})} & \frac{i\omega L}{1 -\omega^{2}L(C_{JJ}+C_{p})}\\
       i \omega C_{g} & 1 
    \end{pmatrix}.
\end{equation}
Here, $L$ is Josephson junction inductance:
\begin{equation}
    L = \frac{\Phi_{0}}{ (2 \pi I_{c})} \left(1 + I_{d}^{2} / (2 I_{c}^{2}) \right).
\end{equation}
The matrix of the unit cell is then given:
\begin{equation}
    M = \begin{pmatrix}
        A & B\\
        C & D 
    \end{pmatrix} =  M_{JJ}^{4}  M_{p}.
\end{equation}
Using the matrix elements, we can approximate the wave vector:\cite{Pozar:882338}
\begin{equation}
\label{eq:plas_dis}
 k \approx  \frac{1}{ndz}\arccos \left(\frac{A+D}{2} \right) = \frac{1}{dx}\arccos \left(\frac{A+D}{2} \right),
\end{equation}
which provides estimated dispersion relation from a single unit cell (see Fig.~\ref{fig:JJ_RPM_pic}(c)).

\section{\label{app:CME3} CME-3}

The CME-3 model describes the evolution of current amplitudes for modes such as signal, idler, pump, sidebands $f_{p+s}$ and $f_{2p-s}$, the pump's second and third harmonics. The case of the pump harmonics falling into the bandgap ($\Im{k}\neq0$) is treated by an approximation, directly including attenuation:\cite{gal2025gaincompressionTWPA} 
\begin{equation}
\frac{dI_{2p,3p}}{dx} = -\delta_{2p,3p} \Im(k_{2p,3p})I_{2p,3p},
\end{equation}
Where $\delta_{2p,3p} $ is the attenuation constant, which we estimated from numerical simulations (see Fig.~\ref{fig:Power_propagation_CME}). In our case, we set $\delta_{2p} =0.00032, \delta_{3p} =0.01 $. This approach avoids divergent behavior of CME for imaginary k-vectors.\cite{Dixon,Dixon_thesis}
Substituting ansatz Eq.~\ref{eq:wave_ansatz} to Eq.~\ref{eq:wave_eq}, the following set of equations is obtained:

\begin{equation}
\label{eq:CME3_I3p}
\begin{aligned}
\frac{d I_{3p}}{dx} =\; &
i\frac{k_{3p} \epsilon}{8} \bigg[2 I_{3p-s} I_{s} e^{ix(k_{3p-s}-k_{3p}+k_s)} + 2 I_{2p} I_{p}  e^{ix(k_{2p}-k_{3p}+k_p)}  + 2 I_{2p+s} I_{p-s}  e^{ix(k_{2p+s}-k_{3p}+k_{p-s})} \\
& + 2 I_{2p-s} I_{p+s}  e^{ix(-k_{3p}+k_{2p-s}+k_{p+s})}
\bigg] \\
& +  i \frac{k_{3p} \xi }{24} \bigg( 6 I_{3p-s} I_{2p}  e^{ix(k_{3p-s}+k_{2p}-k_{3p}-k_{2p-s})} I_{2p-s}^* + 6 I_{3p-s} I_{2p+s}  e^{ix(k_{3p-s}-k_{2p}+k_{2p+s}-k_{3p})} I_{2p}^* \\
& + 6 I_{3p-s} I_{3p}  I_{3p-s}^*  + 6i I_{3p-s} I_{p} k_{3p} \xi e^{ix(k_{3p-s}-k_{3p}-k_{p-s}+k_p)} I_{p-s}^* \\
& + 6 I_{3p-s} I_{p+s} e^{ix(k_{3p-s}-k_{3p}-k_p+k_{p+s})} I_{p}^* + 3 I_{2p}^2 e^{ix(2k_{2p}-k_{3p}-k_p)} I_{p}^* \\
& + 6 I_{2p} I_{2p+s}  e^{ix(k_{2p}+k_{2p+s}-k_{3p}-k_{p+s})} I_{p+s}^*  + 6 I_{2p} I_{3p}  I_{2p}^* \\
& + 6 I_{2p} I_{p-s} I_{s}  e^{ix(k_{2p}-k_{3p}+k_{p-s}+k_s)} + 6 I_{2p} I_{2p-s}  e^{ix(k_{2p}-k_{3p}-k_{p-s}+k_{2p-s})} I_{p-s}^* \\
& + 6 I_{2p} I_{p+s}  e^{ix(k_{2p}-k_{3p}+k_{p+s}-k_s)} I_{s}^*  + 6 I_{2p+s} I_{3p} I_{2p+s}^* \\
& + 6 I_{2p+s} I_{2p-s}  e^{ix(k_{2p+s}-k_{3p}+k_{2p-s}-k_p)} I_{p}^*  + 6 I_{2p+s} I_{p}  e^{ix(k_{2p+s}-k_{3p}+k_p-k_s)} I_{s}^* \\
& + 3 I_{3p}^2  I_{3p}^* + 6 I_{3p} I_{p-s}  I_{p-s}^*  + 6 I_{3p} I_{2p-s}  I_{2p-s}^* + 6 I_{3p} I_{p}  I_{p}^* + 6 I_{3p} I_{p+s}  I_{p+s}^* \\
& + 6 I_{3p} I_{s}  I_{s}^*  + 6 I_{p-s} Ip I_{p+s}  e^{ix(-k_{3p}+k_{p-s}+k_p+k_{p+s})}  + 6 I_{2p-s} I_{p} I_{s}  e^{ix(-k_{3p}+k_{2p-s}+k_p+k_s)} \\
& +  I_{p}^3  e^{ix(-k_{3p}+3k_p)} \bigg)
\end{aligned}
\end{equation}

\begin{equation}
\label{eq:CME3_I3ps}
\begin{aligned} 
\frac{d I_{3p-s}}{dx} =\; &
 i\frac{k_{3p-s} \epsilon}{8}  \bigg[  I_{2p} I_i e^{i x (-k_{3p-s} + k_{2p} + k_i)}
+ I_{3p} e^{i x (-k_{3p-s} + k_{3p} - k_s)} I_s^{*} 
+  I_{ip} I_p e^{i x (-k_{3p-s} + k_{ip} + k_p)}
\bigg] \\
&+ i \frac{k_{3p-s} \xi }{24}  \bigg(3 I_{3p-s}^{2}  I_{3p-s}^* + 6 I_{3p-s} I_{2p} I_{2p}^* + 6 I_{3p-s} I_{2p+s} I_{2p+s}^* + 6 I_{3p-s} I_{3p} I_{3p}^* \\
&+ 6 I_{3p-s} I_{p-s}  I_{p-s}^* + 6 I_{3p-s} I_{2p-s} I_{2p-s}^* + 6 I_{3p-s} I_{p}  I_{p}^* + 6 I_{3p-s} I_{p+s} I_{p+s}^* \\
&+ 6 I_{3p-s} I_{s}  I_{s}^* + 3 I_{2p}^2 e^{ix(-k_{3p-s}+2k_{2p}-k_{p+s})} I_{p+s}^* + 6 I_{2p} I_{3p}  e^{ix(-k_{3p-s}+k_{2p}-k_{2p+s}+k_{3p})} I_{2p+s}^* \\
&+ 6 I_{2p} I_{2p-s}  e^{ix(-k_{3p-s}+k_{2p}+k_{2p-s}-k_p)} I_{p}^* + 6 I_{2p} I_{p}  e^{ix(-k_{3p-s}+k_{2p}+k_p-k_s)} I_{s}^* \\
&+ 6 I_{2p+s} I_{p-s} e^{ix(-k_{3p-s}+k_{2p+s}+k_{p-s}-k_s)} Is^* + 6 I_{2p+s} I_{2p-s}  e^{ix(-k_{3p-s}+k_{2p+s}+k_{2p-s}-k_{p+s})} I_{p+s}^* \\
&+ 6 I_{3p} I_{p-s}  e^{ix(-k_{3p-s}+k_{3p}+k_{p-s}-k_p)} I_{p}^* + 6 I_{3p} I_{2p-s} e^{ix(-k_{3p-s}-k_{2p}+k_{3p}+k_{2p-s})} I_{2p}^* \\
&+ 6 I_{3p} I_{p}  e^{ix(-k_{3p-s}+k_{3p}+k_p-k_{p+s})} I_{p+s}^* + 3 I_{p-s}^2 I_{p+s}  e^{ix(-k_{3p-s}+2k_{p-s}+k_{p+s})} \\
&+ 6 I_{p-s} I_{2p-s} I_{s}  e^{ix(-k_{3p-s}+k_{p-s}+k_{2p-s}+k_s)} + 3 I_{p-s} I_{p}^2  e^{ix(-k_{3p-s}+k_{p-s}+2k_p)} \\
&+ 3 I_{2p-s}^2  e^{ix(-k_{3p-s}-k_{p-s}+2k_{2p-s})} I_{p-s}^* + 6 I_{2p-s} I_{p+s}  e^{ix(-k_{3p-s}+k_{2p-s}+k_{p+s}-k_s)} I_{s}^* \bigg)
\end{aligned}
\end{equation}

\begin{equation}
\label{eq:CME3_I2ps}
\begin{aligned}
\frac{d I_{2p+s}}{dx} =\; &
 i\frac{k_{2p+s} \epsilon}{8} \bigg[2 I_{2p} I_s e^{ix(k_{2p}-k_{2p+s}+k_s)} + 2 I_{3p} e^{ix(-k_{2p+s}+k_{3p}-k_{p-s})} I_{p-s}^* + 2 I_{p} I_{p+s}  e^{ix(-k_{2p+s}+k_p+k_{p+s})}
\bigg] \\
&+i \frac{k_{2p+s} \xi }{24} \bigg( 6 I_{3p-s} I_{2p+s} I_{3p-s}^* + 6 I_{3p-s} I_{p+s}  e^{ix(k_{3p-s}-k_{2p+s}-k_{2p-s}+k_{p+s})} I_{2p-s}^* \\
&+ 6 I_{3p-s} Is  e^{ix(k_{3p-s}-k_{2p+s}-k_{p-s}+k_s)} I_{p-s}^* + 3 I_{2p}^2  e^{ix(2k_{2p}-k_{2p+s}-k_{2p-s})} I_{2p-s}^* \\
&+ 6 I_{2p} I_{2p+s}  I_{2p}^* + 6 I_{2p} I_{3p}  e^{ix(-k_{3p-s}+k_{2p}-k_{2p+s}+k_{3p})} I_{3p-s}^* \\
&+ 6 I_{2p} Ip e^{ix(k_{2p}-k_{2p+s}-k_{p-s}+k_p)} I_{p-s}^* + 6 I_{2p} I_{p+s}  e^{ix(k_{2p}-k_{2p+s}-k_p+k_{p+s})} I_{p}^* \\
&+ 3 I_{2p+s}^2  I_{2p+s}^* + 6 I_{2p+s} I_{3p}  I_{3p}^* + 6 I_{2p+s} I_{p-s} I_{p-s}^* + 6 I_{2p+s} I_{2p-s} I_{2p-s}^* \\
&+ 6 I_{2p+s} I_{p} I_{p}^* + 6 I_{2p+s} I_{p+s} I_{p+s}^* + 6 I_{2p+s} I_{s}  I_{s}^* + 6 I_{3p} I_{p}  e^{ix(-k_{2p+s}+k_{3p}-k_{2p-s}+k_p)} I_{2p-s}^* \\
&+ 6 I_{3p} I_{p+s}  e^{ix(-k_{2p}-k_{2p+s}+k_{3p}+k_{p+s})} I_{2p}^* + 6 I_{3p} Is  e^{ix(-k_{2p+s}+k_{3p}-k_p+k_s)} I_{p}^* \\
&+ 6 I_{p-s} I_{p+s} I_{s} e^{ix(-k_{2p+s}+k_{p-s}+k_{p+s}+k_s)} + 6 I_{2p-s} I_{p+s}  e^{ix(-k_{2p+s}-k_{p-s}+k_{2p-s}+k_{p+s})} I_{p-s}^* \\
&+ 3 I_{2p-s} I_{s}^2  e^{ix(-k_{2p+s}+k_{2p-s}+2k_s)}+ 3 I_{p}^2 Is  e^{ix(-k_{2p+s}+2k_p+k_s)} + 3 I_{p+s}^2 e^{ix(-k_{2p+s}+2k_{p+s}-k_s)} I_{s}^* \bigg)
\end{aligned}
\end{equation}

\begin{equation}
\begin{aligned}
\frac{dI_{2p}}{dx} =& i\frac{k_{2p} \epsilon}{8} \bigg[ 
 I_{3p-s} e^{i x (k_{3p-s} - k_{2p} - k_{p-s})} I_i^* +  I_{2p+s} e^{i x (-k_{2p} + k_{2p+s} - k_s)} I_s^*  +  I_{3p} e^{i x (-k_{2p} + k_{3p} - k_p)} I_p^* \\& + I_{p-s} I_{p+s} e^{i x (-k_{2p} + k_{p-s} + k_{p+s})} +  I_{2p-s} I_s e^{i x (-k_{2p} + k_{2p-s} + k_s)} +  I_p^2 e^{i x (-k_{2p} + 2k_p)} \bigg] \\
 + &i \frac{k_{2p} \xi }{24} \bigg(
 I_{3p-s} I_{2p} I_{3p-s}^* + I_{3p-s} I_{2p+s} e^{i x (k_{3p-s} - k_{2p} + k_{2p+s} - k_{3p})} I_{3p}^* \\
& + I_{3p-s} I_p e^{i x (k_{3p-s} - k_{2p} - k_{2p-s} + k_p)} I_{2p-s}^* + I_{3p-s} I_{p+s} e^{i x (k_{3p-s} - 2k_{2p} + k_{p+s})} I_{2p}^* \\
& + I_{3p-s} I_{s} e^{i x (k_{3p-s} - k_{2p} - k_p + k_s)} I_p^*  + I_{2p}^2 I_{2p}^{*} + I_{2p} I_{2p+s} I_{2p+s}^* + I_{2p} I_{3p} I_{3p}^* \\
& + I_{2p} I_{p-s} I_{p-s}^* + I_{2p} I_{2p-s} I_{2p-s}^* + I_{2p} I_p I_p^* + I_{2p} I_{p+s} I_{p+s}^* + I_{2p} I_s I_s^* \\
& + I_{2p+s} I_{p-s} e^{i x (-k_{2p} + k_{2p+s} + k_{p-s} - k_p)} I_p^* + I_{2p+s} I_{2p-s} e^{i x (-2k_{2p} + k_{2p+s} + k_{2p-s})} I_{2p}^* \\
& + I_{2p+s} I_p e^{i x (-k_{2p} + k_{2p+s} + k_p - k_{p+s})} I_{p+s}^* + I_{3p} I_{p-s} e^{i x (-k_{2p} + k_{3p} + k_{p-s} - k_{2p-s})} I_{2p-s}^* \\
& + I_{3p} I_{2p-s} e^{i x (-k_{3p-s} - k_{2p} + k_{3p} + k_{2p-s})} I_{3p-s}^* + I_{3p} I_p e^{i x (-2k_{2p} + k_{3p} + k_p)} I_{2p}^* \\
& + I_{3p} I_{p+s} e^{i x (-k_{2p} - k_{2p+s} + k_{3p} + k_{p+s})} I_{2p+s}^* + I_{3p} I_s e^{i x (-k_{2p} + k_{3p} - k_{p+s} + k_s)} I_{p+s}^* \\
& + I_{3p} e^{i x (-k_{2p} + k_{3p} - k_{p-s} - k_s)} I_i^* I_s^*  + I_{p-s} I_p I_s e^{i x (-k_{2p} + k_{p-s} + k_p + k_s)} \\
& + I_{2p-s} I_p e^{i x (-k_{2p} - k_{p-s} + k_{2p-s} + k_p)} I_{p-s}^* + I_{2p-s} I_{p+s} e^{i x (-k_{2p} + k_{2p-s} - k_p + k_{p+s})} I_p^* \\
& + I_p I_{p+s} e^{i x (-k_{2p} + k_p + k_{p+s} - k_s)} I_s^* \bigg)
\end{aligned}
\end{equation}

\begin{equation}
\label{eq:CME3_I2p_s}
\begin{aligned}
\frac{d I_{2p-s}}{dx} =\; &
i\frac{k_{2p-s} \epsilon}{8} \bigg[2 I_{3p-s} e^{ix(k_{3p-s}-k_{2p-s}-k_p)} I_{p}^* + 2 I_{2p} e^{ix(k_{2p}-k_{2p-s}-k_s)} I_{s}^* + 2 I_{3p}  e^{ix(k_{3p}-k_{2p-s}-k_{p+s})} I_{p+s}^* \\
&+ 2 I_{p-s} I_{p}  e^{ix(k_{p-s}-k_{2p-s}+k_p)}
\bigg] \\
&+ i \frac{k_{2p-s} \xi }{24} \bigg( 6 I_{3p-s} I_{2p} e^{ix(k_{3p-s}+k_{2p}-k_{3p}-k_{2p-s})} I_{3p}^* + 6 I_{3p-s} I_{p-s} e^{ix(k_{3p-s}+k_{p-s}-2k_{2p-s})} I_{2p-s}^* \\
&+ 6 I_{3p-s} I_{2p-s}  I_{3p-s}^* + 6 I_{3p-s} I_{p} e^{ix(k_{3p-s}-k_{2p}-k_{2p-s}+k_p)} I_{2p}^* \\
&+ 6 I_{3p-s} I_{p+s} e^{ix(k_{3p-s}-k_{2p+s}-k_{2p-s}+k_{p+s})} I_{2p+s}^* + 6 I_{3p-s} I_{s}e^{ix(k_{3p-s}-k_{2p-s}-k_{p+s}+k_s)} I_{p+s}^* \\
&+ 6 I_{3p-s} e^{ix(k_{3p-s}-k_{p-s}-k_{2p-s}-k_s)} I_{p-s}^* I_{s}^* + 3 I_{2p}^2 e^{ix(2k_{2p}-k_{2p+s}-k_{2p-s})} I_{2p+s}^* \\
&+ 6 I_{2p} I_{p-s}  e^{ix(k_{2p}+k_{p-s}-k_{2p-s}-k_p)} I_{p}^* + 6 I_{2p} I_{2p-s}  I_{2p}^* + 6 I_{2p} I_{p}  e^{ix(k_{2p}-k_{2p-s}+k_p-k_{p+s})} I_{p+s}^* \\
&+ 6 I_{2p+s} I_{p-s}  e^{ix(k_{2p+s}+k_{p-s}-k_{2p-s}-k_{p+s})} I_{p+s}^* + 6 I_{2p+s} I_{2p-s}  I_{2p+s}^* \\
&+ 3i I_{2p+s} k_{2p-s} \xi e^{ix(k_{2p+s}-k_{2p-s}-2k_s)} (I_{s}^*)^2 + 6 I_{3p} I_{p-s}  e^{ix(-k_{2p}+k_{3p}+k_{p-s}-k_{2p-s})} I_{2p}^* \\
&+ 6 I_{3p} I_{2p-s} I_{3p}^* + 6 I_{3p} I_{p}  e^{ix(-k_{2p+s}+k_{3p}-k_{2p-s}+k_p)} I_{2p+s}^* \\
&+ 6 I_{3p} e^{ix(k_{3p}-k_{2p-s}-k_p-k_s)} I_{p}^* I_{s}^* + 3 I_{p-s}^2 Is e^{ix(2k_{p-s}-k_{2p-s}+k_s)} \\
&+ 6 I_{p-s} I_{2p-s}  I_{p-s}^* + 6 I_{p-s} I_{p+s}  e^{ix(k_{p-s}-k_{2p-s}+k_{p+s}-k_s)} I_{s}^* + 3 I_{2p-s}^2  I_{2p-s}^* + 6 I_{2p-s} I_{p}  I_{p}^* \\
&+ 6 I_{2p-s} I_{p+s}  I_{p+s}^* + 6 I_{2p-s} I_{s}  I_{s}^* + 3 I_{p}^2  e^{ix(-k_{2p-s}+2k_p-k_s)} I_{s}^* \bigg)
\end{aligned}
\end{equation}

\begin{equation}
\label{eq:CME3_Ips}
\begin{aligned}
\frac{d I_{p+s}}{dx} =\; &
i\frac{k_{p+s} \epsilon}{8} \bigg[ 2 I_{2p}  e^{ix(k_{2p} - k_{p-s} - k_{p+s})} I_{p-s}^{*} + 2 I_{2p+s}  e^{ix(k_{2p+s} - k_{p} - k_{p+s})} Ip^{*} 
+ 2 I_{3p}  e^{ix(k_{3p} - k_{2p-s} - k_{p+s})} I_{2p-s}^{*} \\
&+ 2 I_{p} I_{s} e^{ix(k_{p} - k_{p+s} + k_{s})}
\bigg] \\
&+ i \frac{k_{p+s} \xi }{24}\bigg(6 I_{3p-s} I_{s}  e^{ix(k_{3p-s} - k_{2p-s} - k_{p+s} + k_{s})} I_{2p-s}^{*} + 3 I_{3p-s}  e^{ix(k_{3p-s} - 2k_{p-s} - k_{p+s})} (I_{p-s}^{*})^2 \\
&+ 3 I_{2p}^2  e^{ix(-k_{3p-s} + 2k_{2p} - k_{p+s})} I_{3p-s}^{*} + 6 I_{2p} I_{2p+s}  e^{ix(k_{2p} + k_{2p+s} - k_{3p} - k_{p+s})} I_{3p}^{*} \\
&+ 6 I_{2p} I_{p+s}  I_{2p}^{*} + 6 I_{2p} I_{s}  e^{ix(k_{2p} - k_{p} - k_{p+s} + k_{s})} I_{p}^{*} + 6 I_{2p+s} I_{p-s} e^{ix(k_{2p+s} + k_{p-s} - k_{2p-s} - k_{p+s})} I_{2p-s}^{*} \\
&+ 6 I_{2p+s} I_{2p-s} e^{ix(-k_{3p-s} + k_{2p+s} + k_{2p-s} - k_{p+s})} I_{3p-s}^{*} + 6 I_{2p+s} I_{p+s} I_{2p+s}^{*} + 6 I_{2p+s} I_{s}  e^{ix(k_{2p+s} - 2k_{p+s} + k_{s})} I_{p+s}^{*} \\
&+ 6 I_{2p+s}  e^{ix(k_{2p+s} - k_{p-s} - k_{p+s} - k_{s})} I_{p-s}^{*} I_{s}^{*} + 6 I_{3p} I_{p} e^{ix(-k_{3p-s} + k_{3p} + k_{p} - k_{p+s})} I_{3p-s}^{*} + 6 I_{3p} I_{p+s}  I_{3p}^{*} \\
&+ 6 I_{3p} I_{s}  e^{ix(-k_{2p} + k_{3p} - k_{p+s} + k_{s})} I_{2p}^{*} + 6 I_{3p}  e^{ix(k_{3p} - k_{p-s} - k_{p} - k_{p+s})} I_{p-s}^{*} I_{p}^{*} + 6 I_{p-s} I_{p+s} I_{p-s}^{*}\\ &+ 3 I_{p-s} I_{s}^2 e^{ix(k_{p-s} - k_{p+s} + 2k_{s})} + 6 I_{2p-s} I_{p+s}  I_{2p-s}^{*} + 6 I_{2p-s} I_{s}  e^{ix(-k_{p-s} + k_{2p-s} - k_{p+s} + k_{s})} I_{p-s}^{*} \\
&+ 3 I_{p}^2 e^{ix(-k_{p-s} + 2k_{p} - k_{p+s})} I_{p-s}^{*} + 6 I_{p} I_{p+s}  I_{p}^{*} + 3 I_{p+s}^2  I_{p+s}^{*} + 6 I_{p+s} I_{s}  I_{s}^{*} + 6 I_{3p-s} I_{p+s} I_{3p-s}^{*} \\
&+ 6 I_{2p} I_{p}  e^{ix(k_{2p} - k_{2p-s} + k_{p} - k_{p+s})} I_{2p-s}^{*} + 6 I_{2p+s} I_{p}  e^{ix(-k_{2p} + k_{2p+s} + k_{p} - k_{p+s})} I_{2p}^{*} \bigg)
\end{aligned}
\end{equation}

\begin{equation}
\label{eq:CME3_Ip}
\begin{aligned}
\frac{d I_{p}}{dx} =\;&
 i\frac{k_{p} \epsilon}{8} \bigg[2  I_{3p-s}  e^{i x (k_{3p-s} - k_{2p-s} - k_{p})} I_{2p-s}^{*} 
+ 2  I_{2p}  e^{i x (k_{2p} - 2 k_{p})} I_{p}^{*} 
+ 2  I_{2p+s} e^{i x (k_{2p+s} - k_{p} - k_{p+s})} I_{p+s}^{*} \\
&+ 2  I_{3p}  e^{i x (- k_{2p} + k_{3p} - k_{p})} I_{2p^{*}} 
+ 2  I_{p-s} I_{s} e^{i x (k_{p-s} - k_{p} + k_{s})} 
+ 2  I_{2p-s}  e^{i x (- k_{p-s} + k_{2p-s} - k_{p})} I_{p-s}^{*} \\
&+ 2  I_{p+s}  e^{i x (- k_{p} + k_{p+s} - k_{s})} Is^{*}
\bigg] \\
&+ i \frac{k_{p} \xi }{24} \bigg( 6  I_{3p-s} I_{p}  I_{3p-s}^{*} + 6  I_{3p-s} I_{p+s}  e^{i x (k_{3p-s} - k_{3p} - k_{p} + k_{p+s})} I_{3p}^{*} + 6 I_{3p-s} I_{s}  e^{i x (k_{3p-s} - k_{2p} - k_{p} + k_{s})} I_{2p}^{*} \\
&+ 6  I_{3p-s}  e^{i x (k_{3p-s} - k_{p-s} - 2 k_{p})} I_{p-s}^{*} I_{p}^{*} + 3  I_{2p}^{2}  e^{i x (2 k_{2p} - k_{3p} - k_{p})} I_{3p}^{*} \\
&+ 6  I_{2p} I_{p-s}  e^{i x (k_{2p} + k_{p-s} - k_{2p-s} - k_{p})} I_{2p-s}^{*} + 6  I_{2p} I_{2p-s}  e^{i x (- k_{3p-s} + k_{2p} + k_{2p-s} - k_{p})} I_{3p-s}^{*} + 6  I_{2p} I_{p}  I_{2p}^{*} \\
&+ 6  I_{2p} I_{p+s} e^{i x (k_{2p} - k_{2p+s} - k_{p} + k_{p+s})} I_{2p+s}^{*} + 6  I_{2p} I_{s}  e^{i x (k_{2p} - k_{p} - k_{p+s} + k_{s})} I_{p+s}^{*} \\
&+ 6  I_{2p} e^{i x (k_{2p} - k_{p-s} - k_{p} - k_{s})} I_{p-s}^{*} I_{s}^{*} + 6  I_{2p+s} I_{p-s}  e^{i x (- k_{2p} + k_{2p+s} + k_{p-s} - k_{p})} I_{2p}^{*} \\
&+ 6  I_{2p+s} I_{p} I_{2p+s}^{*} + 6  I_{2p+s}  e^{i x (k_{2p+s} - 2 k_{p} - k_{s})} I_{p}^{*} I_{s}^{*} \\
&+ 6  I_{3p} I_{p-s} e^{i x (- k_{3p-s} + k_{3p} + k_{p-s} - k_{p})} I_{3p-s}^{*} + 6  I_{3p} I_{p}  I_{3p}^{*} + 6 i I_{3p} I_{s} k_{p} \xi e^{i x (- k_{2p+s} + k_{3p} - k_{p} + k_{s})} I_{2p+s}^{*} \\
&+ 3  I_{3p} e^{i x (k_{3p} - 3 k_{p})} (I_{p}^{*})^{2} + 6 I_{3p}  e^{i x (k_{3p} - k_{p-s} - k_{p} - k_{p+s})} I_{p-s}^{*} I_{p+s}^{*} \\
&+ 6  I_{3p}  e^{i x (k_{3p} - k_{2p-s} - k_{p} - k_{s})} I_{2p-s}^{*} I_{s}^{*} + 6  I_{p-s} I_{p} I_{p-s}^{*} + 6  I_{p-s} I_{p+s} e^{i x (k_{p-s} - 2 k_{p} + k_{p+s})} I_{p}^{*} \\
&+ 6 I_{2p-s} I_{p} I_{2p-s}^{*} + 6  I_{2p-s} I_{p+s}  e^{i x (- k_{2p} + k_{2p-s} - k_{p} + k_{p+s})} I_{2p}^{*} + 6 I_{2p-s} Is e^{i x (k_{2p-s} - 2 k_{p} + k_{s})} I_{p}^{*} \\
&+ 3  I_{p}^{2}  I_{p}^{*} + 6 I_{p} I_{p+s} I_{p+s}^{*} + 6 I_{p} I_{s} I_{s}^{*} + 6  I_{2p+s} I_{2p-s} e^{i x (k_{2p+s} - k_{3p} + k_{2p-s} - k_{p})} I_{3p}^{*} \bigg)
\end{aligned}
\end{equation}

\begin{equation}
\label{eq:CME3_Ip_s}
\begin{aligned}
\frac{d I_{p-s}}{dx} =\;&
 i\frac{k_{p-s} \epsilon}{8} \Bigg[
2  I_{3p-s}  k_{p-s} e^{i x (k_{3p-s} - k_{2p} - k_{p-s})} I_{2p}^{*} + 2  I_{2p}  k_{p-s} e^{i x (k_{2p} - k_{p-s} - k_{p+s})} I_{p+s}^{*} \\
& + 2 i I_{3p} k_{p-s} e^{i x (- k_{2p+s} + k_{3p} - k_{p-s})} I_{2p+s}^{*} + 2  I_{2p-s}  k_{p-s} e^{i x (- k_{p-s} + k_{2p-s} - k_{p})} I_p^{*} + 2  I_{p}  k_{p-s} e^{i x (- k_{p-s} + k_{p} - k_{s})} I_{s}^{*}
\Bigg] \\
& + i \frac{k_{p-s} \xi }{24}  \bigg( 6  I_{3p-s} I_{p-s} I_{3p-s}^{*} + 6  I_{3p-s} I_{p}  e^{i x (k_{3p-s} - k_{3p} - k_{p-s} + k_{p})} I_{3p}^{*} + 6  I_{3p-s} I_{s} e^{i x (k_{3p-s} - k_{2p+s} - k_{p-s} + k_{s})} I_{2p+s}^{*} \\
& + 6  I_{3p-s} e^{i x (k_{3p-s} - 2 k_{p-s} - k_{p+s})} I_{p-s}^{*} I_{p+s}^{*} + 3  I_{3p-s}  e^{i x (k_{3p-s} - k_{p-s} - 2 k_{p})} (I_{p}^{*})^{2} \\
& + 6 I_{3p-s}  e^{i x (k_{3p-s} - k_{p-s} - k_{2p-s} - k_{s})} I_{2p-s}^{*} I_{s}^{*} + 6  I_{2p} I_{p-s}  I_{2p}^{*} + 6  I_{2p} I_{2p-s}  e^{i x (k_{2p} - k_{3p} - k_{p-s} + k_{2p-s})} I_{3p}^{*} \\
& + 6  I_{2p} I_{p}  e^{i x (k_{2p} - k_{2p+s} - k_{p-s} + k_{p})} I_{2p+s}^{*} + 6  I_{2p}  e^{i x (k_{2p} - k_{p-s} - k_{p} - k_{s})} I_{p}^{*} I_{s}^{*} \\
& + 6  I_{2p+s} I_{p-s}  I_{2p+s}^{*} + 6  I_{2p+s}  e^{i x (k_{2p+s} - k_{p-s} - k_{p+s} - k_{s})} I_{p+s}^{*} I_{s}^{*} + 6  I_{3p} I_{p-s}  I_{3p}^{*} \\
& + 6 I_{3p}  e^{i x (- k_{2p} + k_{3p} - k_{p-s} - k_{s})} I_{2p}^{*} I_{s}^{*} + 6  I_{3p}  e^{i x (k_{3p} - k_{p-s} - k_{p} - k_{p+s})} I_{p}^{*} I_{p+s}^{*} \\
& + 3  I_{p-s}^{2}  I_{p-s}^{*} + 6  I_{p-s} I_{2p-s}  I_{2p-s}^{*} + 6  I_{p-s} I_{p}  I_{p}^{*} + 6  I_{p-s} I_{p+s}  I_{p+s}^{*} + 6  I_{p-s} I_{s}  I_{s}^{*} \\
& + 3  I_{2p-s}^{2}  e^{i x (- k_{3p-s} - k_{p-s} + 2 k_{2p-s})} I_{3p-s}^{*} + 6  I_{2p-s} I_{p}  e^{i x (- k_{2p} - k_{p-s} + k_{2p-s} + k_{p})} I_{2p}^{*} \\
& + 6  I_{2p-s} I_{p+s}  e^{i x (- k_{2p+s} - k_{p-s} + k_{2p-s} + k_{p+s})} I_{2p+s}^{*} + 6  I_{2p-s} I_{s}  e^{i x (- k_{p-s} + k_{2p-s} - k_{p+s} + k_{s})} I_{p+s}^{*} \\
& + 6  I_{2p-s}  e^{i x (- 2 k_{p-s} + k_{2p-s} - k_{s})} I_{p-s}^{*} I_{s}^{*} + 3  I_{p}^{2}  e^{i x (- k_{p-s} + 2 k_{p} - k_{p+s})} I_{p+s}^{*} \\ &+ 3  I_{p+s} e^{i x (- k_{p-s} + k_{p+s} - 2 k_{s})} (I_{s}^{*})^{2} \bigg)
\end{aligned}
\end{equation}

\begin{equation}
\label{eq:CME3_Is}
\begin{aligned}
\frac{d I_{s}}{dx} =\; &
i\frac{k_{s} \epsilon}{8}  \left[
2 i I_{2p} \epsilon k_{s} e^{i x (k_{2p} - k_{2p-s} - k_{s})} I_{2p-s}^{*}
+ 2 i I_{2p+s} \epsilon k_{s} e^{i x (-k_{2p} + k_{2p+s} - k_{s})} I_{2p}^{*}
+ 2 i I_{3p} \epsilon k_{s} e^{i x (-k_{3p-s} + k_{3p} - k_{s})} I_{3p-s}^{*}
\right. \\
& \left.
+ 2 i I_{p} \epsilon k_{s} e^{i x (-k_{p-s} + k_{p} - k_{s})} I_{p-s}^{*}
+ 2 i I_{p+s} \epsilon k_{s} e^{i x (-k_{p} + k_{p+s} - k_{s})} I_{p}^{*}
\right]
\\
& +i \frac{k_{s} \xi }{24} \bigg( 6  I_{3p-s} I_{s}  I_{3p-s}^{*}
+ 6  I_{3p-s}  e^{i x (k_{3p-s} - k_{p-s} - k_{2p-s} - k_{s})} I_{p-s}^{*} I_{2p-s}^{*}
+ 6  I_{2p} I_{p}  e^{i x (-k_{3p-s} + k_{2p} + k_{p} - k_{s})} I_{3p-s}^{*} \\
& + 6  I_{2p} I_{p+s}  e^{i x (k_{2p} - k_{3p} + k_{p+s} - k_{s})} I_{3p}^{*}
+ 6  I_{2p} I_{s}  I_{2p}^{*}
+ 6 I_{2p}  e^{i x (k_{2p} - k_{p-s} - k_{p} - k_{s})} I_{p-s}^{*} I_{p}^{*} \\
& + 6  I_{2p+s} I_{p-s} e^{i x (-k_{3p-s} + k_{2p+s} + k_{p-s} - k_{s})} I_{3p-s}^{*}
+ 6  I_{2p+s} I_{p}  e^{i x (k_{2p+s} - k_{3p} + k_{p} - k_{s})} I_{3p}^{*}
+ 6  I_{2p+s} I_{s}  I_{2p+s}^{*} \\
& + 6  I_{2p+s}  e^{i x (k_{2p+s} - k_{2p-s} - 2k_{s})} I_{2p-s}^{*} I_{s}^{*}
+ 3  I_{2p+s} e^{i x (k_{2p+s} - 2k_{p} - k_{s})} I_{p}^{*2}
+ 6  I_{2p+s} e^{i x (k_{2p+s} - k_{p-s} - k_{p+s} - k_{s})} I_{p-s}^{*} I_{p+s}^{*} \\
& + 6  I_{3p} I_{s} I_{3p}^{*}
+ 6  I_{3p}  e^{i x (-k_{2p} + k_{3p} - k_{p-s} - k_{s})} I_{2p}^{*} I_{p-s}^{*}
+ 6  I_{3p} e^{i x (k_{3p} - k_{2p-s} - k_{p} - k_{s})} I_{2p-s}^{*} I_{p}^{*} \\
& + 6  I_{p-s} I_{p+s}  e^{i x (k_{p-s} - k_{2p-s} + k_{p+s} - k_{s})} I_{2p-s}^{*}
+ 6  I_{p-s} I_{s}  I_{p-s}^{*}
+ 6  I_{2p-s} I_{p+s}  e^{i x (-k_{3p-s} + k_{2p-s} + k_{p+s} - k_{s})} I_{3p-s}^{*} \\
& + 6  I_{2p-s} I_{s} I_{2p-s}^{*}
+ 3  I_{2p-s} e^{i x (-2k_{p-s} + k_{2p-s} - k_{s})} I_{p-s}^{*2}
+ 3  I_{p}^{2}  e^{i x (-k_{2p-s} + 2k_{p} - k_{s})} I_{2p-s}^{*} \\
& + 6  I_{p} I_{p+s} e^{i x (-k_{2p} + k_{p} + k_{p+s} - k_{s})} I_{2p}^{*}
+ 6  I_{p} I_{s}  I_{p}^{*}
+ 3  I_{p+s}^{2}  e^{i x (-k_{2p+s} + 2k_{p+s} - k_{s})} I_{2p+s}^{*} \\
& + 6  I_{p+s} I_{s}  I_{p+s}^{*}
+ 6 I_{p+s} e^{i x (-k_{p-s} + k_{p+s} - 2k_{s})} I_{p-s}^{*} I_{s}^{*}
+ 3  I_{s}^{2} I_{s}^{*} \bigg)
\end{aligned}
\end{equation}

To solve this system we tested different methods in solve\_ivp function of SciPy library in Python:\cite{Solve_ivp_cite} Explicit Runge-Kutta method of order 5(4), 3(2), 8; \cite{RK23,RK45,RK8} and implicit multi-step variable-order (1 to 5) method based on a backward differentiation formula for the derivative approximation (BDF). \cite{BDF} We ensured that all orders of the Runge-Kutta method give the same solution, which is close to BDF method solution. It is essential to fix the minimal step during the simulation, which is the length of the Unit Cell $dx = 5 dz = 25~\mu \textrm{m} $, considering smaller steps have no physical meaning. Note that Runge-Kutta methods in solve\_ivp use adaptive simulation steps, and it could choose simulation steps smaller than the length of a segment of the waveguide corresponding to a single JJ, which leads to divergent solutions. To fix this, we set the relative and the absolute tolerance to large numbers rtol = 1e+90, atol = 1e+90, and set max\_step = first\_step = $25~\mu \textrm{m} $. For the BDF method, we set rtol = 1e-6 and atol = 1e-9.
The solution of this system by the Explicit Runge-Kutta method of order 3(2) is depicted in Fig. \ref{fig:Power_propagation_CME}. The simulation code used is available online.\cite{CMT_github}

\section{The dominant processes behind THG mediated gain within CME}
\label{app:dominant}
A simplified version of coupled-mode theory was derived by retaining only the terms essential for capturing the connection between third harmonic strength and signal gain:
\begin{equation}
\label{eq:dp3p}
I_p^\prime = i\frac{k_p \xi}{24}\left(  I_p \left(3\left|I_p\right|^2  + 6\left|I_{3p}\right|^2\right)\right), \quad
I_{3p}^\prime = i\frac{k_{3p} \xi}{24}\left(  I_{3p} \left( 6\left|I_p\right|^2  + 3\left|I_{3p}\right|^2 \right)\right)
\end{equation}
\begin{equation}
\label{eq:ds}
I_s^\prime = i\frac{k_s\epsilon}{8}\left(  2I_{p} I_{p-s}^\ast e^{i\left(k_{p}-k_{p-s}-k_s\right)x}  +  2I_{p+s} I_p^\ast e^{i\left(k_{p+s}-k_p-k_s\right)x}\right) +i\frac{k_s \xi}{24}\left(  I_s \left(6\left|I_p\right|^2  + 6\left|I_{3p}\right|^2\right)\right)
\end{equation}
\begin{equation}
\label{eq:di}
I_{p-s}^\prime = i\frac{k_{p-s}\epsilon}{8}\left(  2I_{p} I_s^\ast e^{i\left(k_{p}-k_s-k_{p-s}\right)x}  \right) +i\frac{k_{p-s} \xi}{24}\left(  I_{p-s} \left(6\left|I_p\right|^2  + 6\left|I_{3p}\right|^2\right)\right)
\end{equation}
\begin{equation}
\label{eq:dpps}
I_{p+s}^\prime = i\frac{k_{p+s} \epsilon}{8} \left(  2I_sI_p e^{i\left(k_s+k_p-k_{p+s}\right)x} 
+ 2I_{3p}I_{2p-s}^\ast e^{i\left(k_{3p}-k_{2p-s}-k_{p+s}\right)x}\right)+i\frac{k_{p+s} \xi}{24}\left(  I_{p+s} \left(6\left|I_p\right|^2  + 6\left|I_{3p}\right|^2\right)\right)
\end{equation}
\begin{equation}
\label{eq:ddpms}
I_{2p-s}^\prime = i\frac{k_{2p-s} \epsilon}{8} \left(2 I_{3p} I_{p+s}^\ast e^{i\left(k_{3p}-k_{p+s}-k_{2p-s}\right)x}\right)+i\frac{k_{2p-s} \xi}{24}\left(  I_{2p-s} \left(6\left|I_p\right|^2  + 6\left|I_{3p}\right|^2\right)\right)
\end{equation}

Utilizing undepleted pump and third hamornic approximation, Eqs. (\ref{eq:dp3p}) yields
\begin{equation}
\label{eq:p_3p_sub}
I_p = I_p(0)\exp\left(i\frac{k_p\xi}{24}\left(3\left|I_{p}(0)\right|^2+6\left|I_{3p}(0)\right|^2\right)x\right), \quad I_{3p} = I_{3p}(0)\exp\left(i\frac{k_{3p}\xi}{24}\left(6\left|I_{p}(0)\right|^2+3\left|I_{3p}(0)\right|^2\right)x\right)
\end{equation}
Making use of the following substitutions
\begin{equation}
Y_s = I_s \exp\left(-i\frac{k_s\xi}{24}\left(6\left|I_{p}(0)\right|^2+6\left|I_{3p}(0)\right|^2\right)x\right) \exp\left(i\frac{1}{2}\frac{k_p\xi}{24}3\left|I_{p}(0)\right|^2x\right)
\end{equation}
\begin{equation}
Y_{p-s} = I_{p-s}^\ast \exp\left(i\frac{k_{p-s}\xi}{24}\left(6\left|I_{p}(0)\right|^2+6\left|I_{3p}(0)\right|^2\right)x\right) \exp\left(-i\frac{1}{2}\frac{k_p\xi}{24}3\left|I_{p}(0)\right|^2x\right)
\end{equation}
\begin{equation}
Y_{p+s} = I_{p+s} \exp\left(-i\frac{k_{p+s}\xi}{24}\left(6\left|I_{p}(0)\right|^2+6\left|I_{3p}(0)\right|^2\right)x\right) \exp\left(i\frac{3}{2}\frac{k_p\xi}{24}3\left|I_{p}(0)\right|^2x\right)
\end{equation}
\begin{equation}
\label{eq:Y2ps}
Y_{2p-s} = I_{2p-s}^\ast \exp\left(i\frac{k_{2p-s}\xi}{24}\left(6\left|I_{p}(0)\right|^2+6\left|I_{3p}(0)\right|^2\right)x\right) \exp\left(i\left(\frac{3}{2}\frac{k_p\xi}{24}3\left|I_{p}(0)\right|^2-\left(\theta+\frac{k_{3p}\xi}{24}3\left|I_{3p}(0)\right|^2\right)\right)x\right)
\end{equation}
the system Eqs. (\ref{eq:dp3p}-\ref{eq:ddpms}) can be expressed in a matrix form as presented in the main text.

\section{Robustness of Gain and THG Under Critical Current Variations} 
\label{ch:Parameter_variation}

\begin{figure}[h]

    \includegraphics[width=8.6cm]{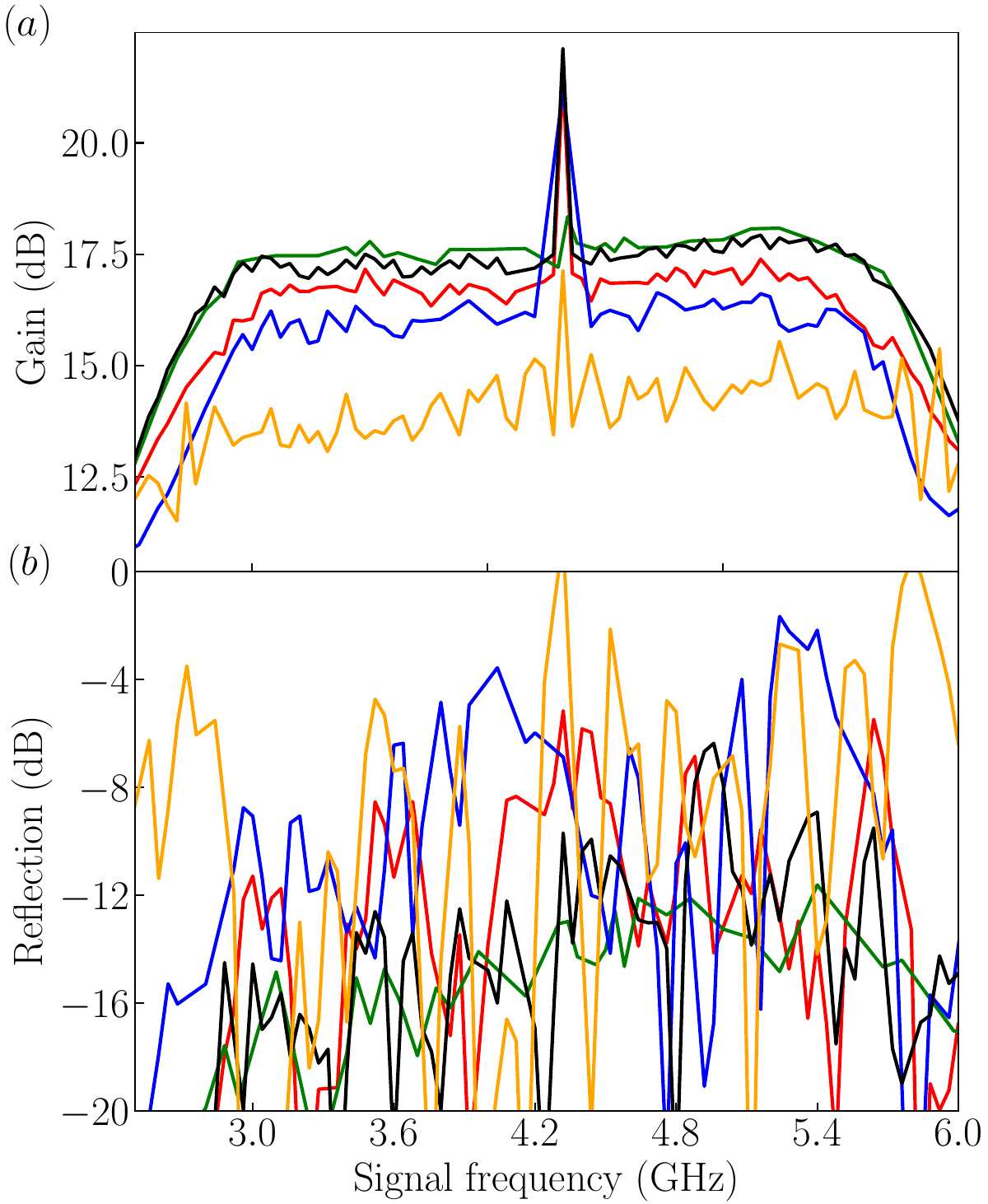}
    \caption{\label{fig:Gain_spread_comparing}  Median Gain (a) and reflections (b) simulated in JoSIM for $\sigma = 0\%$ (green), $2.5\%$ (black), $5\%$ (red), $7.5\%$ (blue), $10\%$ (orange). Parameters of the simulation are: $f_{p} = 8.64 \textrm{ GHz}, I_{p} = 1.6 ~\mu\textrm{A},  I_{s} = 0.01~\mu\textrm{A}, I_{d} = 0.8~\mu\textrm{A}$.} 
\end{figure}

\begin{figure}[h]

    \includegraphics[width=8.6cm]{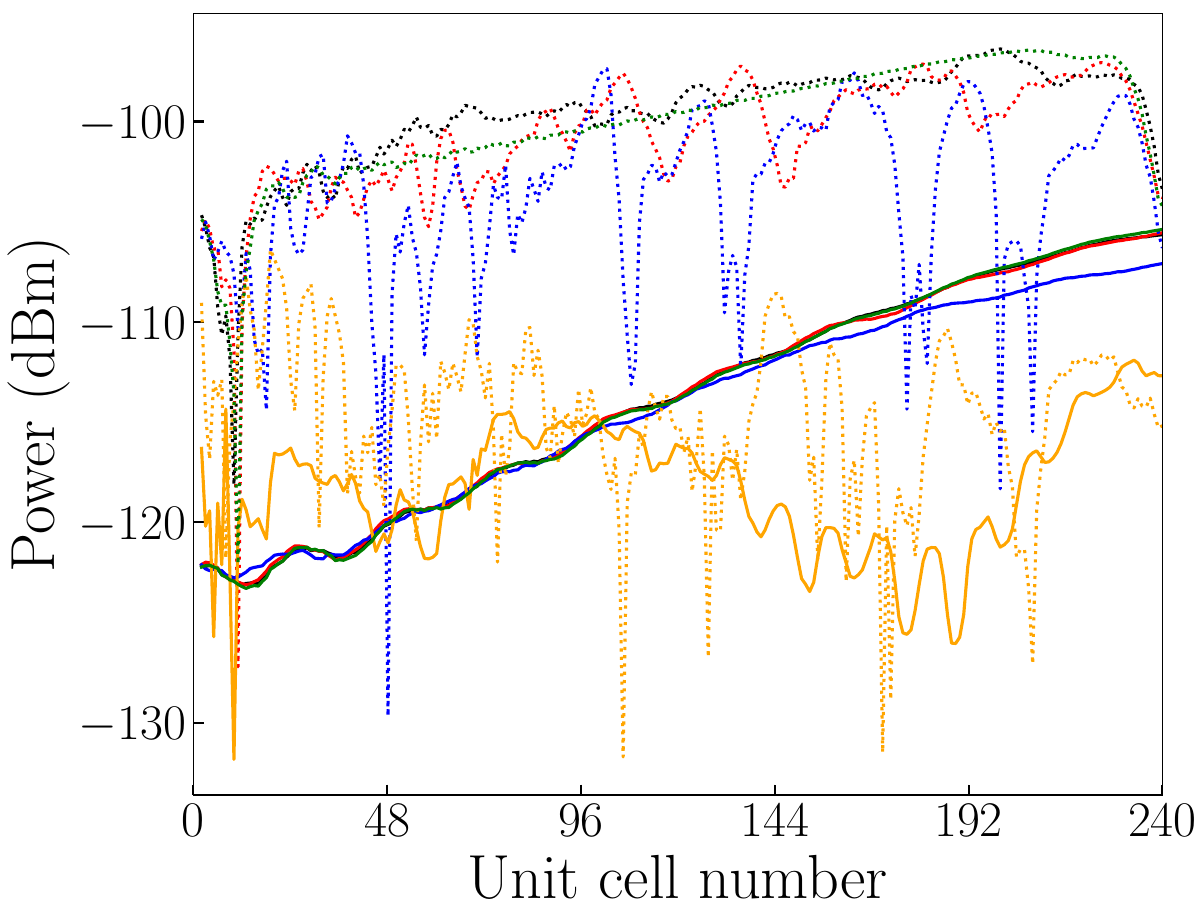}
    \caption{\label{fig:Power_flow_with_spread}  Power flow of propagating plane waves for signal and third harmonic of the pump simulated in JoSIM, each line corresponding to a different standard deviation: $\sigma = 0\%$ (green), $2.5\%$ (black), $5\%$ (red), $7.5\%$ (blue), $10\%$ (orange). Parameters of the simulation are: $f_{p} = 8.64 \textrm{ GHz}, f_{s} = 4.8 \textrm{ GHz}, I_{p} = 1.6 ~\mu\textrm{A},  I_{s} = 0.01~\mu\textrm{A}, I_{d} = 0.8~\mu\textrm{A}$.} 
\end{figure}

To verify the robustness of gain and third harmonic generation (THG) against technological imperfections, we considered spatial variations in the junction critical current using a Gaussian distribution, following the approach used in previous works. \cite{Peng_FM,Kissling2023,Peatain2021,Samolov} As a result, we plot the median gain profile  (Fig.\ref{fig:Gain_spread_comparing}) and the median power flow for the signal and third harmonic generation (THG) (Fig.\ref{fig:Power_flow_with_spread}), corresponding to the distribution that lies in the middle of the ensemble. These results are shown for a fixed mean value of  2 \textrm{uA} and varying standard deviations:  $\sigma = 0\%,5\%,7.5\%,10\%$.

The critical-current variation mainly affects the reflection similarly to that in Ref.\onlinecite{Peng_FM} and THG. 

\newpage

\bibliographystyle{aipnum4-1}
\bibliography{references}

\begin{thebibliography}{50}%
\makeatletter
\providecommand \@ifxundefined [1]{%
 \@ifx{#1\undefined}
}%
\providecommand \@ifnum [1]{%
 \ifnum #1\expandafter \@firstoftwo
 \else \expandafter \@secondoftwo
 \fi
}%
\providecommand \@ifx [1]{%
 \ifx #1\expandafter \@firstoftwo
 \else \expandafter \@secondoftwo
 \fi
}%
\providecommand \natexlab [1]{#1}%
\providecommand \enquote  [1]{``#1''}%
\providecommand \bibnamefont  [1]{#1}%
\providecommand \bibfnamefont [1]{#1}%
\providecommand \citenamefont [1]{#1}%
\providecommand \href@noop [0]{\@secondoftwo}%
\providecommand \href [0]{\begingroup \@sanitize@url \@href}%
\providecommand \@href[1]{\@@startlink{#1}\@@href}%
\providecommand \@@href[1]{\endgroup#1\@@endlink}%
\providecommand \@sanitize@url [0]{\catcode `\\12\catcode `\$12\catcode `\&12\catcode `\#12\catcode `\^12\catcode `\_12\catcode `\%12\relax}%
\providecommand \@@startlink[1]{}%
\providecommand \@@endlink[0]{}%
\providecommand \url  [0]{\begingroup\@sanitize@url \@url }%
\providecommand \@url [1]{\endgroup\@href {#1}{\urlprefix }}%
\providecommand \urlprefix  [0]{URL }%
\providecommand \Eprint [0]{\href }%
\providecommand \doibase [0]{http://dx.doi.org/}%
\providecommand \selectlanguage [0]{\@gobble}%
\providecommand \bibinfo  [0]{\@secondoftwo}%
\providecommand \bibfield  [0]{\@secondoftwo}%
\providecommand \translation [1]{[#1]}%
\providecommand \BibitemOpen [0]{}%
\providecommand \bibitemStop [0]{}%
\providecommand \bibitemNoStop [0]{.\EOS\space}%
\providecommand \EOS [0]{\spacefactor3000\relax}%
\providecommand \BibitemShut  [1]{\csname bibitem#1\endcsname}%
\let\auto@bib@innerbib\@empty
\bibitem [{\citenamefont {Devoret}\ and\ \citenamefont {Schoelkopf}(2013)}]{Devoret}%
  \BibitemOpen
  \bibfield  {author} {\bibinfo {author} {\bibfnamefont {M.~H.}\ \bibnamefont {Devoret}}\ and\ \bibinfo {author} {\bibfnamefont {R.~J.}\ \bibnamefont {Schoelkopf}},\ }\href@noop {} {\bibfield  {journal} {\bibinfo  {journal} {Science}\ }\textbf {\bibinfo {volume} {339}},\ \bibinfo {pages} {1169} (\bibinfo {year} {2013})}\BibitemShut {NoStop}%
\bibitem [{\citenamefont {Citro}, \citenamefont {Guarcello},\ and\ \citenamefont {Pagano}(2024)}]{Pagano_book}%
  \BibitemOpen
  \bibfield  {author} {\bibinfo {author} {\bibfnamefont {R.}~\bibnamefont {Citro}}, \bibinfo {author} {\bibfnamefont {C.}~\bibnamefont {Guarcello}}, \ and\ \bibinfo {author} {\bibfnamefont {S.}~\bibnamefont {Pagano}},\ }\enquote {\bibinfo {title} {Josephson junctions, superconducting circuits, and qubit for quantum technologies},}\ in\ \href {\doibase 10.1007/978-3-031-55657-9_1} {\emph {\bibinfo {booktitle} {New Trends and Platforms for Quantum Technologies}}},\ \bibinfo {editor} {edited by\ \bibinfo {editor} {\bibfnamefont {R.}~\bibnamefont {Aguado}}, \bibinfo {editor} {\bibfnamefont {R.}~\bibnamefont {Citro}}, \bibinfo {editor} {\bibfnamefont {M.}~\bibnamefont {Lewenstein}}, \ and\ \bibinfo {editor} {\bibfnamefont {M.}~\bibnamefont {Stern}}}\ (\bibinfo  {publisher} {Springer Nature Switzerland},\ \bibinfo {address} {Cham},\ \bibinfo {year} {2024})\ pp.\ \bibinfo {pages} {1--59}\BibitemShut {NoStop}%
\bibitem [{\citenamefont {Macklin}\ \emph {et~al.}(2015)\citenamefont {Macklin}, \citenamefont {O’Brien}, \citenamefont {Hover}, \citenamefont {Schwartz}, \citenamefont {Bolkhovsky}, \citenamefont {Zhang}, \citenamefont {Oliver},\ and\ \citenamefont {Siddiqi}}]{Macklin}%
  \BibitemOpen
  \bibfield  {author} {\bibinfo {author} {\bibfnamefont {C.}~\bibnamefont {Macklin}}, \bibinfo {author} {\bibfnamefont {K.}~\bibnamefont {O’Brien}}, \bibinfo {author} {\bibfnamefont {D.}~\bibnamefont {Hover}}, \bibinfo {author} {\bibfnamefont {M.~E.}\ \bibnamefont {Schwartz}}, \bibinfo {author} {\bibfnamefont {V.}~\bibnamefont {Bolkhovsky}}, \bibinfo {author} {\bibfnamefont {X.}~\bibnamefont {Zhang}}, \bibinfo {author} {\bibfnamefont {W.~D.}\ \bibnamefont {Oliver}}, \ and\ \bibinfo {author} {\bibfnamefont {I.}~\bibnamefont {Siddiqi}},\ }\href {\doibase 10.1126/science.aaa8525} {\bibfield  {journal} {\bibinfo  {journal} {Science}\ }\textbf {\bibinfo {volume} {350}},\ \bibinfo {pages} {307} (\bibinfo {year} {2015})}\BibitemShut {NoStop}%
\bibitem [{\citenamefont {Pagano}\ \emph {et~al.}(2022)\citenamefont {Pagano}, \citenamefont {Barone}, \citenamefont {Borghesi}, \citenamefont {Chung}, \citenamefont {Carapella}, \citenamefont {Caricato}, \citenamefont {Carusotto}, \citenamefont {Cian}, \citenamefont {Gioacchino}, \citenamefont {Enrico}, \citenamefont {Falferi}, \citenamefont {Fasolo}, \citenamefont {Faverzani}, \citenamefont {Ferri}, \citenamefont {Filatrella}, \citenamefont {Gatti}, \citenamefont {Giachero}, \citenamefont {Giubertoni}, \citenamefont {Greco}, \citenamefont {Kutlu}, \citenamefont {Leo}, \citenamefont {Ligi}, \citenamefont {Maccarrone}, \citenamefont {Margesin}, \citenamefont {Maruccio}, \citenamefont {Matlashov}, \citenamefont {Mauro}, \citenamefont {Mezzena}, \citenamefont {Monteduro}, \citenamefont {Nucciotti}, \citenamefont {Oberto}, \citenamefont {Pierro}, \citenamefont {Piersanti}, \citenamefont {Rajteri}, \citenamefont {Rettaroli}, \citenamefont {Rizzato}, \citenamefont {Semertzidis}, \citenamefont {Uchaikin},\ and\
  \citenamefont {Vinante}}]{Pagano2}%
  \BibitemOpen
  \bibfield  {author} {\bibinfo {author} {\bibfnamefont {S.}~\bibnamefont {Pagano}}, \bibinfo {author} {\bibfnamefont {C.}~\bibnamefont {Barone}}, \bibinfo {author} {\bibfnamefont {M.}~\bibnamefont {Borghesi}}, \bibinfo {author} {\bibfnamefont {W.}~\bibnamefont {Chung}}, \bibinfo {author} {\bibfnamefont {G.}~\bibnamefont {Carapella}}, \bibinfo {author} {\bibfnamefont {A.~P.}\ \bibnamefont {Caricato}}, \bibinfo {author} {\bibfnamefont {I.}~\bibnamefont {Carusotto}}, \bibinfo {author} {\bibfnamefont {A.}~\bibnamefont {Cian}}, \bibinfo {author} {\bibfnamefont {D.~D.}\ \bibnamefont {Gioacchino}}, \bibinfo {author} {\bibfnamefont {E.}~\bibnamefont {Enrico}}, \bibinfo {author} {\bibfnamefont {P.}~\bibnamefont {Falferi}}, \bibinfo {author} {\bibfnamefont {L.}~\bibnamefont {Fasolo}}, \bibinfo {author} {\bibfnamefont {M.}~\bibnamefont {Faverzani}}, \bibinfo {author} {\bibfnamefont {E.}~\bibnamefont {Ferri}}, \bibinfo {author} {\bibfnamefont {G.}~\bibnamefont {Filatrella}}, \bibinfo {author} {\bibfnamefont
  {C.}~\bibnamefont {Gatti}}, \bibinfo {author} {\bibfnamefont {A.}~\bibnamefont {Giachero}}, \bibinfo {author} {\bibfnamefont {D.}~\bibnamefont {Giubertoni}}, \bibinfo {author} {\bibfnamefont {A.}~\bibnamefont {Greco}}, \bibinfo {author} {\bibfnamefont {C.}~\bibnamefont {Kutlu}}, \bibinfo {author} {\bibfnamefont {A.}~\bibnamefont {Leo}}, \bibinfo {author} {\bibfnamefont {C.}~\bibnamefont {Ligi}}, \bibinfo {author} {\bibfnamefont {G.}~\bibnamefont {Maccarrone}}, \bibinfo {author} {\bibfnamefont {B.}~\bibnamefont {Margesin}}, \bibinfo {author} {\bibfnamefont {G.}~\bibnamefont {Maruccio}}, \bibinfo {author} {\bibfnamefont {A.}~\bibnamefont {Matlashov}}, \bibinfo {author} {\bibfnamefont {C.}~\bibnamefont {Mauro}}, \bibinfo {author} {\bibfnamefont {R.}~\bibnamefont {Mezzena}}, \bibinfo {author} {\bibfnamefont {A.~G.}\ \bibnamefont {Monteduro}}, \bibinfo {author} {\bibfnamefont {A.}~\bibnamefont {Nucciotti}}, \bibinfo {author} {\bibfnamefont {L.}~\bibnamefont {Oberto}}, \bibinfo {author} {\bibfnamefont
  {V.}~\bibnamefont {Pierro}}, \bibinfo {author} {\bibfnamefont {L.}~\bibnamefont {Piersanti}}, \bibinfo {author} {\bibfnamefont {M.}~\bibnamefont {Rajteri}}, \bibinfo {author} {\bibfnamefont {A.}~\bibnamefont {Rettaroli}}, \bibinfo {author} {\bibfnamefont {S.}~\bibnamefont {Rizzato}}, \bibinfo {author} {\bibfnamefont {Y.~K.}\ \bibnamefont {Semertzidis}}, \bibinfo {author} {\bibfnamefont {S.}~\bibnamefont {Uchaikin}}, \ and\ \bibinfo {author} {\bibfnamefont {A.}~\bibnamefont {Vinante}},\ }\href {\doibase 10.1109/TASC.2022.3145782} {\bibfield  {journal} {\bibinfo  {journal} {IEEE Transactions on Applied Superconductivity}\ }\textbf {\bibinfo {volume} {32}},\ \bibinfo {pages} {1} (\bibinfo {year} {2022})}\BibitemShut {NoStop}%
\bibitem [{\citenamefont {Planat}\ \emph {et~al.}(2020)\citenamefont {Planat}, \citenamefont {Ranadive}, \citenamefont {Dassonneville}, \citenamefont {Puertas~Mart\'{\i}nez}, \citenamefont {L\'eger}, \citenamefont {Naud}, \citenamefont {Buisson}, \citenamefont {Hasch-Guichard}, \citenamefont {Basko},\ and\ \citenamefont {Roch}}]{Planat2019}%
  \BibitemOpen
  \bibfield  {author} {\bibinfo {author} {\bibfnamefont {L.}~\bibnamefont {Planat}}, \bibinfo {author} {\bibfnamefont {A.}~\bibnamefont {Ranadive}}, \bibinfo {author} {\bibfnamefont {R.}~\bibnamefont {Dassonneville}}, \bibinfo {author} {\bibfnamefont {J.}~\bibnamefont {Puertas~Mart\'{\i}nez}}, \bibinfo {author} {\bibfnamefont {S.}~\bibnamefont {L\'eger}}, \bibinfo {author} {\bibfnamefont {C.}~\bibnamefont {Naud}}, \bibinfo {author} {\bibfnamefont {O.}~\bibnamefont {Buisson}}, \bibinfo {author} {\bibfnamefont {W.}~\bibnamefont {Hasch-Guichard}}, \bibinfo {author} {\bibfnamefont {D.~M.}\ \bibnamefont {Basko}}, \ and\ \bibinfo {author} {\bibfnamefont {N.}~\bibnamefont {Roch}},\ }\href {\doibase 10.1103/PhysRevX.10.021021} {\bibfield  {journal} {\bibinfo  {journal} {Phys. Rev. X}\ }\textbf {\bibinfo {volume} {10}},\ \bibinfo {pages} {021021} (\bibinfo {year} {2020})}\BibitemShut {NoStop}%
\bibitem [{\citenamefont {Bell}\ and\ \citenamefont {Samolov}(2015)}]{Samolov}%
  \BibitemOpen
  \bibfield  {author} {\bibinfo {author} {\bibfnamefont {M.~T.}\ \bibnamefont {Bell}}\ and\ \bibinfo {author} {\bibfnamefont {A.}~\bibnamefont {Samolov}},\ }\href {\doibase 10.1103/PhysRevApplied.4.024014} {\bibfield  {journal} {\bibinfo  {journal} {Phys. Rev. Appl.}\ }\textbf {\bibinfo {volume} {4}},\ \bibinfo {pages} {024014} (\bibinfo {year} {2015})}\BibitemShut {NoStop}%
\bibitem [{\citenamefont {Ranadive}\ \emph {et~al.}(2022)\citenamefont {Ranadive}, \citenamefont {Esposito}, \citenamefont {Planat}, \citenamefont {Bonet}, \citenamefont {Naud}, \citenamefont {Buisson}, \citenamefont {Guichard},\ and\ \citenamefont {Roch}}]{Ranadive_2022}%
  \BibitemOpen
  \bibfield  {author} {\bibinfo {author} {\bibfnamefont {A.}~\bibnamefont {Ranadive}}, \bibinfo {author} {\bibfnamefont {M.}~\bibnamefont {Esposito}}, \bibinfo {author} {\bibfnamefont {L.}~\bibnamefont {Planat}}, \bibinfo {author} {\bibfnamefont {E.}~\bibnamefont {Bonet}}, \bibinfo {author} {\bibfnamefont {C.}~\bibnamefont {Naud}}, \bibinfo {author} {\bibfnamefont {O.}~\bibnamefont {Buisson}}, \bibinfo {author} {\bibfnamefont {W.}~\bibnamefont {Guichard}}, \ and\ \bibinfo {author} {\bibfnamefont {N.}~\bibnamefont {Roch}},\ }\href@noop {} {\bibfield  {journal} {\bibinfo  {journal} {Nature Communications}\ }\textbf {\bibinfo {volume} {13}} (\bibinfo {year} {2022})}\BibitemShut {NoStop}%
\bibitem [{\citenamefont {Zorin}(2016)}]{Zorin2016}%
  \BibitemOpen
  \bibfield  {author} {\bibinfo {author} {\bibfnamefont {A.~B.}\ \bibnamefont {Zorin}},\ }\href {\doibase 10.1103/PhysRevApplied.6.034006} {\bibfield  {journal} {\bibinfo  {journal} {Phys. Rev. Appl.}\ }\textbf {\bibinfo {volume} {6}},\ \bibinfo {pages} {034006} (\bibinfo {year} {2016})}\BibitemShut {NoStop}%
\bibitem [{\citenamefont {Zorin}\ \emph {et~al.}(2017)\citenamefont {Zorin}, \citenamefont {Khabipov}, \citenamefont {Dietel},\ and\ \citenamefont {Dolata}}]{Zorin2017}%
  \BibitemOpen
  \bibfield  {author} {\bibinfo {author} {\bibfnamefont {A.~B.}\ \bibnamefont {Zorin}}, \bibinfo {author} {\bibfnamefont {M.}~\bibnamefont {Khabipov}}, \bibinfo {author} {\bibfnamefont {J.}~\bibnamefont {Dietel}}, \ and\ \bibinfo {author} {\bibfnamefont {R.}~\bibnamefont {Dolata}},\ }in\ \href {\doibase 10.1109/ISEC.2017.8314196} {\emph {\bibinfo {booktitle} {2017 16th International Superconductive Electronics Conference (ISEC)}}}\ (\bibinfo {year} {2017})\ pp.\ \bibinfo {pages} {1--3}\BibitemShut {NoStop}%
\bibitem [{\citenamefont {Dixon}\ \emph {et~al.}(2020)\citenamefont {Dixon}, \citenamefont {Dunstan}, \citenamefont {Long}, \citenamefont {Williams}, \citenamefont {Meeson},\ and\ \citenamefont {Shelly}}]{Dixon}%
  \BibitemOpen
  \bibfield  {author} {\bibinfo {author} {\bibfnamefont {T.}~\bibnamefont {Dixon}}, \bibinfo {author} {\bibfnamefont {J.}~\bibnamefont {Dunstan}}, \bibinfo {author} {\bibfnamefont {G.}~\bibnamefont {Long}}, \bibinfo {author} {\bibfnamefont {J.}~\bibnamefont {Williams}}, \bibinfo {author} {\bibfnamefont {P.}~\bibnamefont {Meeson}}, \ and\ \bibinfo {author} {\bibfnamefont {C.}~\bibnamefont {Shelly}},\ }\href {\doibase 10.1103/PhysRevApplied.14.034058} {\bibfield  {journal} {\bibinfo  {journal} {Phys. Rev. Appl.}\ }\textbf {\bibinfo {volume} {14}},\ \bibinfo {pages} {034058} (\bibinfo {year} {2020})}\BibitemShut {NoStop}%
\bibitem [{\citenamefont {Gaydamachenko}\ \emph {et~al.}(2022)\citenamefont {Gaydamachenko}, \citenamefont {Kissling}, \citenamefont {Dolata},\ and\ \citenamefont {Zorin}}]{Gaydamachenko2022}%
  \BibitemOpen
  \bibfield  {author} {\bibinfo {author} {\bibfnamefont {V.}~\bibnamefont {Gaydamachenko}}, \bibinfo {author} {\bibfnamefont {C.}~\bibnamefont {Kissling}}, \bibinfo {author} {\bibfnamefont {R.}~\bibnamefont {Dolata}}, \ and\ \bibinfo {author} {\bibfnamefont {A.~B.}\ \bibnamefont {Zorin}},\ }\href {\doibase 10.1063/5.0111111} {\bibfield  {journal} {\bibinfo  {journal} {Journal of Applied Physics}\ }\textbf {\bibinfo {volume} {132}},\ \bibinfo {pages} {154401} (\bibinfo {year} {2022})}\BibitemShut {NoStop}%
\bibitem [{\citenamefont {Kissling}\ \emph {et~al.}(2023)\citenamefont {Kissling}, \citenamefont {Gaydamachenko}, \citenamefont {Kaap}, \citenamefont {Khabipov}, \citenamefont {Dolata}, \citenamefont {Zorin},\ and\ \citenamefont {Grünhaupt}}]{Kissling2023}%
  \BibitemOpen
  \bibfield  {author} {\bibinfo {author} {\bibfnamefont {C.}~\bibnamefont {Kissling}}, \bibinfo {author} {\bibfnamefont {V.}~\bibnamefont {Gaydamachenko}}, \bibinfo {author} {\bibfnamefont {F.}~\bibnamefont {Kaap}}, \bibinfo {author} {\bibfnamefont {M.}~\bibnamefont {Khabipov}}, \bibinfo {author} {\bibfnamefont {R.}~\bibnamefont {Dolata}}, \bibinfo {author} {\bibfnamefont {A.~B.}\ \bibnamefont {Zorin}}, \ and\ \bibinfo {author} {\bibfnamefont {L.}~\bibnamefont {Grünhaupt}},\ }\href {\doibase 10.1109/TASC.2023.3242927} {\bibfield  {journal} {\bibinfo  {journal} {IEEE Transactions on Applied Superconductivity}\ }\textbf {\bibinfo {volume} {33}},\ \bibinfo {pages} {1} (\bibinfo {year} {2023})}\BibitemShut {NoStop}%
\bibitem [{\citenamefont {Renberg~Nilsson}\ \emph {et~al.}(2024)\citenamefont {Renberg~Nilsson}, \citenamefont {Shiri}, \citenamefont {Rehammar}, \citenamefont {Fadavi~Roudsari},\ and\ \citenamefont {Delsing}}]{Nilsson2023}%
  \BibitemOpen
  \bibfield  {author} {\bibinfo {author} {\bibfnamefont {H.}~\bibnamefont {Renberg~Nilsson}}, \bibinfo {author} {\bibfnamefont {D.}~\bibnamefont {Shiri}}, \bibinfo {author} {\bibfnamefont {R.}~\bibnamefont {Rehammar}}, \bibinfo {author} {\bibfnamefont {A.}~\bibnamefont {Fadavi~Roudsari}}, \ and\ \bibinfo {author} {\bibfnamefont {P.}~\bibnamefont {Delsing}},\ }\href@noop {} {\bibfield  {journal} {\bibinfo  {journal} {Phys. Rev. Appl.}\ }\textbf {\bibinfo {volume} {21}},\ \bibinfo {pages} {064062} (\bibinfo {year} {2024})}\BibitemShut {NoStop}%
\bibitem [{\citenamefont {Peatain}\ \emph {et~al.}(2023)\citenamefont {Peatain}, \citenamefont {Meeson}, \citenamefont {Williams}, \citenamefont {Kafanov},\ and\ \citenamefont {Pashkin}}]{Peatain2021}%
  \BibitemOpen
  \bibfield  {author} {\bibinfo {author} {\bibfnamefont {T.}~\bibnamefont {Peatain}, \bibfnamefont {Searbhán Oand~Dixon}}, \bibinfo {author} {\bibfnamefont {P.~J.}\ \bibnamefont {Meeson}}, \bibinfo {author} {\bibfnamefont {J.}~\bibnamefont {Williams}}, \bibinfo {author} {\bibfnamefont {S.}~\bibnamefont {Kafanov}}, \ and\ \bibinfo {author} {\bibfnamefont {Y.~A.}\ \bibnamefont {Pashkin}},\ }\href {\doibase 10.1088/1361-6668/acba4e} {\bibfield  {journal} {\bibinfo  {journal} {Superconductor Science and Technology}\ }\textbf {\bibinfo {volume} {36}},\ \bibinfo {pages} {045017} (\bibinfo {year} {2023})}\BibitemShut {NoStop}%
\bibitem [{\citenamefont {O'Brien}\ \emph {et~al.}(2014)\citenamefont {O'Brien}, \citenamefont {Macklin}, \citenamefont {Siddiqi},\ and\ \citenamefont {Zhang}}]{OBrien2014}%
  \BibitemOpen
  \bibfield  {author} {\bibinfo {author} {\bibfnamefont {K.}~\bibnamefont {O'Brien}}, \bibinfo {author} {\bibfnamefont {C.}~\bibnamefont {Macklin}}, \bibinfo {author} {\bibfnamefont {I.}~\bibnamefont {Siddiqi}}, \ and\ \bibinfo {author} {\bibfnamefont {X.}~\bibnamefont {Zhang}},\ }\href@noop {} {\bibfield  {journal} {\bibinfo  {journal} {Phys. Rev. Lett. 113, 157001 (2014)}\ }\textbf {\bibinfo {volume} {113}},\ \bibinfo {pages} {157001} (\bibinfo {year} {2014})}\BibitemShut {NoStop}%
\bibitem [{\citenamefont {White}\ \emph {et~al.}(2015)\citenamefont {White}, \citenamefont {Mutus}, \citenamefont {Hoi}, \citenamefont {Barends}, \citenamefont {Campbell}, \citenamefont {Chen}, \citenamefont {Chen}, \citenamefont {Chiaro}, \citenamefont {Dunsworth}, \citenamefont {Jeffrey}, \citenamefont {Kelly}, \citenamefont {Megrant}, \citenamefont {Neill}, \citenamefont {O'Malley}, \citenamefont {Roushan}, \citenamefont {Sank}, \citenamefont {Vainsencher}, \citenamefont {Wenner}, \citenamefont {Chaudhuri}, \citenamefont {Gao},\ and\ \citenamefont {Martinis}}]{White2015}%
  \BibitemOpen
  \bibfield  {author} {\bibinfo {author} {\bibfnamefont {T.~C.}\ \bibnamefont {White}}, \bibinfo {author} {\bibfnamefont {J.~Y.}\ \bibnamefont {Mutus}}, \bibinfo {author} {\bibfnamefont {I.-C.}\ \bibnamefont {Hoi}}, \bibinfo {author} {\bibfnamefont {R.}~\bibnamefont {Barends}}, \bibinfo {author} {\bibfnamefont {B.}~\bibnamefont {Campbell}}, \bibinfo {author} {\bibfnamefont {Y.}~\bibnamefont {Chen}}, \bibinfo {author} {\bibfnamefont {Z.}~\bibnamefont {Chen}}, \bibinfo {author} {\bibfnamefont {B.}~\bibnamefont {Chiaro}}, \bibinfo {author} {\bibfnamefont {A.}~\bibnamefont {Dunsworth}}, \bibinfo {author} {\bibfnamefont {E.}~\bibnamefont {Jeffrey}}, \bibinfo {author} {\bibfnamefont {J.}~\bibnamefont {Kelly}}, \bibinfo {author} {\bibfnamefont {A.}~\bibnamefont {Megrant}}, \bibinfo {author} {\bibfnamefont {C.}~\bibnamefont {Neill}}, \bibinfo {author} {\bibfnamefont {P.~J.~J.}\ \bibnamefont {O'Malley}}, \bibinfo {author} {\bibfnamefont {P.}~\bibnamefont {Roushan}}, \bibinfo {author} {\bibfnamefont {D.}~\bibnamefont
  {Sank}}, \bibinfo {author} {\bibfnamefont {A.}~\bibnamefont {Vainsencher}}, \bibinfo {author} {\bibfnamefont {J.}~\bibnamefont {Wenner}}, \bibinfo {author} {\bibfnamefont {S.}~\bibnamefont {Chaudhuri}}, \bibinfo {author} {\bibfnamefont {J.}~\bibnamefont {Gao}}, \ and\ \bibinfo {author} {\bibfnamefont {J.~M.}\ \bibnamefont {Martinis}},\ }\href {\doibase 10.1063/1.4922348} {\bibfield  {journal} {\bibinfo  {journal} {Applied Physics Letters}\ }\textbf {\bibinfo {volume} {106}},\ \bibinfo {pages} {242601} (\bibinfo {year} {2015})}\BibitemShut {NoStop}%
\bibitem [{\citenamefont {Peng}\ \emph {et~al.}(2022{\natexlab{a}})\citenamefont {Peng}, \citenamefont {Poore}, \citenamefont {Krantz}, \citenamefont {Root},\ and\ \citenamefont {O’Brien}}]{Peng_2022}%
  \BibitemOpen
  \bibfield  {author} {\bibinfo {author} {\bibfnamefont {K.}~\bibnamefont {Peng}}, \bibinfo {author} {\bibfnamefont {R.}~\bibnamefont {Poore}}, \bibinfo {author} {\bibfnamefont {P.}~\bibnamefont {Krantz}}, \bibinfo {author} {\bibfnamefont {D.~E.}\ \bibnamefont {Root}}, \ and\ \bibinfo {author} {\bibfnamefont {K.~P.}\ \bibnamefont {O’Brien}},\ }in\ \href {\doibase 10.1109/qce53715.2022.00054} {\emph {\bibinfo {booktitle} {2022 IEEE International Conference on Quantum Computing and Engineering (QCE)}}}\ (\bibinfo  {publisher} {IEEE},\ \bibinfo {year} {2022})\BibitemShut {NoStop}%
\bibitem [{\citenamefont {Peng}\ \emph {et~al.}(2022{\natexlab{b}})\citenamefont {Peng}, \citenamefont {Naghiloo}, \citenamefont {Wang}, \citenamefont {Cunningham}, \citenamefont {Ye},\ and\ \citenamefont {O’Brien}}]{Peng_FM}%
  \BibitemOpen
  \bibfield  {author} {\bibinfo {author} {\bibfnamefont {K.}~\bibnamefont {Peng}}, \bibinfo {author} {\bibfnamefont {M.}~\bibnamefont {Naghiloo}}, \bibinfo {author} {\bibfnamefont {J.}~\bibnamefont {Wang}}, \bibinfo {author} {\bibfnamefont {G.~D.}\ \bibnamefont {Cunningham}}, \bibinfo {author} {\bibfnamefont {Y.}~\bibnamefont {Ye}}, \ and\ \bibinfo {author} {\bibfnamefont {K.~P.}\ \bibnamefont {O’Brien}},\ }\href {http://dx.doi.org/10.1103/PRXQuantum.3.020306} {\bibfield  {journal} {\bibinfo  {journal} {PRX Quantum}\ }\textbf {\bibinfo {volume} {3}} (\bibinfo {year} {2022}{\natexlab{b}})}\BibitemShut {NoStop}%
\bibitem [{\citenamefont {Yusupov}\ \emph {et~al.}(2022)\citenamefont {Yusupov}, \citenamefont {Filippenko}, \citenamefont {Bazulin}, \citenamefont {Kolotinskiy}, \citenamefont {Tarasov}, \citenamefont {Goldobin}, \citenamefont {Koshelets},\ and\ \citenamefont {Kornev}}]{BiSQUID}%
  \BibitemOpen
  \bibfield  {author} {\bibinfo {author} {\bibfnamefont {R.~A.}\ \bibnamefont {Yusupov}}, \bibinfo {author} {\bibfnamefont {L.~V.}\ \bibnamefont {Filippenko}}, \bibinfo {author} {\bibfnamefont {D.~E.}\ \bibnamefont {Bazulin}}, \bibinfo {author} {\bibfnamefont {N.~V.}\ \bibnamefont {Kolotinskiy}}, \bibinfo {author} {\bibfnamefont {M.~A.}\ \bibnamefont {Tarasov}}, \bibinfo {author} {\bibfnamefont {E.}~\bibnamefont {Goldobin}}, \bibinfo {author} {\bibfnamefont {V.~P.}\ \bibnamefont {Koshelets}}, \ and\ \bibinfo {author} {\bibfnamefont {V.~K.}\ \bibnamefont {Kornev}},\ }\href {\doibase 10.1109/TASC.2021.3131134} {\bibfield  {journal} {\bibinfo  {journal} {IEEE Transactions on Applied Superconductivity}\ }\textbf {\bibinfo {volume} {32}},\ \bibinfo {pages} {1} (\bibinfo {year} {2022})}\BibitemShut {NoStop}%
\bibitem [{\citenamefont {Yurke}\ \emph {et~al.}(1989)\citenamefont {Yurke}, \citenamefont {Corruccini}, \citenamefont {Kaminsky}, \citenamefont {Rupp}, \citenamefont {Smith}, \citenamefont {Silver}, \citenamefont {Simon},\ and\ \citenamefont {Whittaker}}]{PhysRevA.39.2519}%
  \BibitemOpen
  \bibfield  {author} {\bibinfo {author} {\bibfnamefont {B.}~\bibnamefont {Yurke}}, \bibinfo {author} {\bibfnamefont {L.~R.}\ \bibnamefont {Corruccini}}, \bibinfo {author} {\bibfnamefont {P.~G.}\ \bibnamefont {Kaminsky}}, \bibinfo {author} {\bibfnamefont {L.~W.}\ \bibnamefont {Rupp}}, \bibinfo {author} {\bibfnamefont {A.~D.}\ \bibnamefont {Smith}}, \bibinfo {author} {\bibfnamefont {A.~H.}\ \bibnamefont {Silver}}, \bibinfo {author} {\bibfnamefont {R.~W.}\ \bibnamefont {Simon}}, \ and\ \bibinfo {author} {\bibfnamefont {E.~A.}\ \bibnamefont {Whittaker}},\ }\href {\doibase 10.1103/PhysRevA.39.2519} {\bibfield  {journal} {\bibinfo  {journal} {Phys. Rev. A}\ }\textbf {\bibinfo {volume} {39}},\ \bibinfo {pages} {2519} (\bibinfo {year} {1989})}\BibitemShut {NoStop}%
\bibitem [{\citenamefont {Sweeny}\ and\ \citenamefont {Mahler}(1985)}]{First_JTWPA}%
  \BibitemOpen
  \bibfield  {author} {\bibinfo {author} {\bibfnamefont {M.}~\bibnamefont {Sweeny}}\ and\ \bibinfo {author} {\bibfnamefont {R.}~\bibnamefont {Mahler}},\ }\href {\doibase 10.1109/TMAG.1985.1063777} {\bibfield  {journal} {\bibinfo  {journal} {IEEE Transactions on Magnetics}\ }\textbf {\bibinfo {volume} {21}},\ \bibinfo {pages} {654} (\bibinfo {year} {1985})}\BibitemShut {NoStop}%
\bibitem [{\citenamefont {Yaakobi}\ \emph {et~al.}(2013)\citenamefont {Yaakobi}, \citenamefont {Friedland}, \citenamefont {Macklin},\ and\ \citenamefont {Siddiqi}}]{PhysRevB.87.144301}%
  \BibitemOpen
  \bibfield  {author} {\bibinfo {author} {\bibfnamefont {O.}~\bibnamefont {Yaakobi}}, \bibinfo {author} {\bibfnamefont {L.}~\bibnamefont {Friedland}}, \bibinfo {author} {\bibfnamefont {C.}~\bibnamefont {Macklin}}, \ and\ \bibinfo {author} {\bibfnamefont {I.}~\bibnamefont {Siddiqi}},\ }\href@noop {} {\bibfield  {journal} {\bibinfo  {journal} {Phys. Rev. B}\ }\textbf {\bibinfo {volume} {87}},\ \bibinfo {pages} {144301} (\bibinfo {year} {2013})}\BibitemShut {NoStop}%
\bibitem [{\citenamefont {Kow}\ and\ \citenamefont {Bell}(2025)}]{kow2025traveling}%
  \BibitemOpen
  \bibfield  {author} {\bibinfo {author} {\bibfnamefont {C.}~\bibnamefont {Kow}}\ and\ \bibinfo {author} {\bibfnamefont {M.}~\bibnamefont {Bell}},\ }\href@noop {} {\bibfield  {journal} {\bibinfo  {journal} {arXiv preprint arXiv:2505.04059}\ } (\bibinfo {year} {2025})}\BibitemShut {NoStop}%
\bibitem [{\citenamefont {Kern}\ \emph {et~al.}(2023)\citenamefont {Kern}, \citenamefont {Neilinger}, \citenamefont {Il'ichev}, \citenamefont {Sultanov}, \citenamefont {Schmelz}, \citenamefont {Linzen}, \citenamefont {Kunert}, \citenamefont {Oelsner}, \citenamefont {Stolz}, \citenamefont {Danilov}, \citenamefont {Mahashabde}, \citenamefont {Jayaraman}, \citenamefont {Antonov}, \citenamefont {Kubatkin},\ and\ \citenamefont {Grajcar}}]{Kern2022}%
  \BibitemOpen
  \bibfield  {author} {\bibinfo {author} {\bibfnamefont {S.}~\bibnamefont {Kern}}, \bibinfo {author} {\bibfnamefont {P.}~\bibnamefont {Neilinger}}, \bibinfo {author} {\bibfnamefont {E.}~\bibnamefont {Il'ichev}}, \bibinfo {author} {\bibfnamefont {A.}~\bibnamefont {Sultanov}}, \bibinfo {author} {\bibfnamefont {M.}~\bibnamefont {Schmelz}}, \bibinfo {author} {\bibfnamefont {S.}~\bibnamefont {Linzen}}, \bibinfo {author} {\bibfnamefont {J.}~\bibnamefont {Kunert}}, \bibinfo {author} {\bibfnamefont {G.}~\bibnamefont {Oelsner}}, \bibinfo {author} {\bibfnamefont {R.}~\bibnamefont {Stolz}}, \bibinfo {author} {\bibfnamefont {A.}~\bibnamefont {Danilov}}, \bibinfo {author} {\bibfnamefont {S.}~\bibnamefont {Mahashabde}}, \bibinfo {author} {\bibfnamefont {A.}~\bibnamefont {Jayaraman}}, \bibinfo {author} {\bibfnamefont {V.}~\bibnamefont {Antonov}}, \bibinfo {author} {\bibfnamefont {S.}~\bibnamefont {Kubatkin}}, \ and\ \bibinfo {author} {\bibfnamefont {M.}~\bibnamefont {Grajcar}},\ }\href@noop {} {\bibfield  {journal}
  {\bibinfo  {journal} {Phys. Rev. B}\ }\textbf {\bibinfo {volume} {107}},\ \bibinfo {pages} {174520} (\bibinfo {year} {2023})}\BibitemShut {NoStop}%
\bibitem [{\citenamefont {Aumentado}(2020)}]{9134828}%
  \BibitemOpen
  \bibfield  {author} {\bibinfo {author} {\bibfnamefont {J.}~\bibnamefont {Aumentado}},\ }\href {\doibase 10.1109/MMM.2020.2993476} {\bibfield  {journal} {\bibinfo  {journal} {IEEE Microwave Magazine}\ }\textbf {\bibinfo {volume} {21}},\ \bibinfo {pages} {45} (\bibinfo {year} {2020})}\BibitemShut {NoStop}%
\bibitem [{\citenamefont {Eom}\ \emph {et~al.}(2012)\citenamefont {Eom}, \citenamefont {Day}, \citenamefont {Leduc},\ and\ \citenamefont {Zmuidzinas}}]{Eom2012}%
  \BibitemOpen
  \bibfield  {author} {\bibinfo {author} {\bibfnamefont {B.~H.}\ \bibnamefont {Eom}}, \bibinfo {author} {\bibfnamefont {P.~K.}\ \bibnamefont {Day}}, \bibinfo {author} {\bibfnamefont {H.~G.}\ \bibnamefont {Leduc}}, \ and\ \bibinfo {author} {\bibfnamefont {J.}~\bibnamefont {Zmuidzinas}},\ }\href {\doibase 10.48550/ARXIV.1201.2392} {\  (\bibinfo {year} {2012}),\ 10.48550/ARXIV.1201.2392},\ \Eprint {http://arxiv.org/abs/1201.2392} {arXiv:1201.2392 [cond-mat.supr-con]} \BibitemShut {NoStop}%
\bibitem [{\citenamefont {Dixon}(2022)}]{Dixon_thesis}%
  \BibitemOpen
  \bibfield  {author} {\bibinfo {author} {\bibfnamefont {T.}~\bibnamefont {Dixon}},\ }\emph {\bibinfo {title} {Resolving Many Mode Processes in Josephson Travelling Wave Parametric Amplifiers}},\ \href@noop {} {Ph.D. thesis},\ \bibinfo  {school} {Royal Holloway, University of London} (\bibinfo {year} {2022})\BibitemShut {NoStop}%
\bibitem [{\citenamefont {Levochkina}\ \emph {et~al.}(2024)\citenamefont {Levochkina}, \citenamefont {Ahmad}, \citenamefont {Mastrovito}, \citenamefont {Chatterjee}, \citenamefont {Serpico}, \citenamefont {Di~Palma}, \citenamefont {Ferroiuolo}, \citenamefont {Satariano}, \citenamefont {Darvehi}, \citenamefont {Ranadive}, \citenamefont {Cappelli}, \citenamefont {Le~Gal}, \citenamefont {Planat}, \citenamefont {Montemurro}, \citenamefont {Massarotti}, \citenamefont {Tafuri}, \citenamefont {Roch}, \citenamefont {Pepe},\ and\ \citenamefont {Esposito}}]{levochkinaTHG}%
  \BibitemOpen
  \bibfield  {author} {\bibinfo {author} {\bibfnamefont {A.~Y.}\ \bibnamefont {Levochkina}}, \bibinfo {author} {\bibfnamefont {H.~G.}\ \bibnamefont {Ahmad}}, \bibinfo {author} {\bibfnamefont {P.}~\bibnamefont {Mastrovito}}, \bibinfo {author} {\bibfnamefont {I.}~\bibnamefont {Chatterjee}}, \bibinfo {author} {\bibfnamefont {G.}~\bibnamefont {Serpico}}, \bibinfo {author} {\bibfnamefont {L.}~\bibnamefont {Di~Palma}}, \bibinfo {author} {\bibfnamefont {R.}~\bibnamefont {Ferroiuolo}}, \bibinfo {author} {\bibfnamefont {R.}~\bibnamefont {Satariano}}, \bibinfo {author} {\bibfnamefont {P.}~\bibnamefont {Darvehi}}, \bibinfo {author} {\bibfnamefont {A.}~\bibnamefont {Ranadive}}, \bibinfo {author} {\bibfnamefont {G.}~\bibnamefont {Cappelli}}, \bibinfo {author} {\bibfnamefont {G.}~\bibnamefont {Le~Gal}}, \bibinfo {author} {\bibfnamefont {L.}~\bibnamefont {Planat}}, \bibinfo {author} {\bibfnamefont {D.}~\bibnamefont {Montemurro}}, \bibinfo {author} {\bibfnamefont {D.}~\bibnamefont {Massarotti}}, \bibinfo {author}
  {\bibfnamefont {F.}~\bibnamefont {Tafuri}}, \bibinfo {author} {\bibfnamefont {N.}~\bibnamefont {Roch}}, \bibinfo {author} {\bibfnamefont {G.~P.}\ \bibnamefont {Pepe}}, \ and\ \bibinfo {author} {\bibfnamefont {M.}~\bibnamefont {Esposito}},\ }\href {\doibase 10.1088/1361-6668/ad8314} {\bibfield  {journal} {\bibinfo  {journal} {Superconductor Science and Technology}\ }\textbf {\bibinfo {volume} {37}},\ \bibinfo {pages} {115021} (\bibinfo {year} {2024})}\BibitemShut {NoStop}%
\bibitem [{\citenamefont {Fadavi~Roudsari}\ \emph {et~al.}(2023)\citenamefont {Fadavi~Roudsari}, \citenamefont {Shiri}, \citenamefont {Renberg~Nilsson}, \citenamefont {Tancredi}, \citenamefont {Osman}, \citenamefont {Svensson}, \citenamefont {Kudra}, \citenamefont {Rommel}, \citenamefont {Bylander}, \citenamefont {Shumeiko},\ and\ \citenamefont {Delsing}}]{Nilsom_Photonic_crystall}%
  \BibitemOpen
  \bibfield  {author} {\bibinfo {author} {\bibfnamefont {A.}~\bibnamefont {Fadavi~Roudsari}}, \bibinfo {author} {\bibfnamefont {D.}~\bibnamefont {Shiri}}, \bibinfo {author} {\bibfnamefont {H.}~\bibnamefont {Renberg~Nilsson}}, \bibinfo {author} {\bibfnamefont {G.}~\bibnamefont {Tancredi}}, \bibinfo {author} {\bibfnamefont {A.}~\bibnamefont {Osman}}, \bibinfo {author} {\bibfnamefont {I.-M.}\ \bibnamefont {Svensson}}, \bibinfo {author} {\bibfnamefont {M.}~\bibnamefont {Kudra}}, \bibinfo {author} {\bibfnamefont {M.}~\bibnamefont {Rommel}}, \bibinfo {author} {\bibfnamefont {J.}~\bibnamefont {Bylander}}, \bibinfo {author} {\bibfnamefont {V.}~\bibnamefont {Shumeiko}}, \ and\ \bibinfo {author} {\bibfnamefont {P.}~\bibnamefont {Delsing}},\ }\href {\doibase 10.1063/5.0127690} {\bibfield  {journal} {\bibinfo  {journal} {Applied Physics Letters}\ }\textbf {\bibinfo {volume} {122}},\ \bibinfo {pages} {052601} (\bibinfo {year} {2023})}\BibitemShut {NoStop}%
\bibitem [{\citenamefont {Renberg~Nilsson}\ \emph {et~al.}(2023)\citenamefont {Renberg~Nilsson}, \citenamefont {Fadavi}, \citenamefont {Shiri}, \citenamefont {Delsing},\ and\ \citenamefont {Shumeiko}}]{RenbergNilsson2023}%
  \BibitemOpen
  \bibfield  {author} {\bibinfo {author} {\bibfnamefont {H.}~\bibnamefont {Renberg~Nilsson}}, \bibinfo {author} {\bibfnamefont {A.}~\bibnamefont {Fadavi}}, \bibinfo {author} {\bibfnamefont {D.}~\bibnamefont {Shiri}}, \bibinfo {author} {\bibfnamefont {P.}~\bibnamefont {Delsing}}, \ and\ \bibinfo {author} {\bibfnamefont {V.}~\bibnamefont {Shumeiko}},\ }\href@noop {} {\bibfield  {journal} {\bibinfo  {journal} {Physical Review Applied}\ }\textbf {\bibinfo {volume} {19}} (\bibinfo {year} {2023})}\BibitemShut {NoStop}%
\bibitem [{\citenamefont {Rizvanov}\ \emph {et~al.}(2024)\citenamefont {Rizvanov}, \citenamefont {Kern}, \citenamefont {Neilinger},\ and\ \citenamefont {Grajcar}}]{PTWPA}%
  \BibitemOpen
  \bibfield  {author} {\bibinfo {author} {\bibfnamefont {E.}~\bibnamefont {Rizvanov}}, \bibinfo {author} {\bibfnamefont {S.}~\bibnamefont {Kern}}, \bibinfo {author} {\bibfnamefont {P.}~\bibnamefont {Neilinger}}, \ and\ \bibinfo {author} {\bibfnamefont {M.}~\bibnamefont {Grajcar}},\ }\href@noop {} {\bibfield  {journal} {\bibinfo  {journal} {Journal of Applied Physics}\ }\textbf {\bibinfo {volume} {136}},\ \bibinfo {pages} {174401} (\bibinfo {year} {2024})}\BibitemShut {NoStop}%
\bibitem [{\citenamefont {Guarcello}\ \emph {et~al.}(2025)\citenamefont {Guarcello}, \citenamefont {Barone}, \citenamefont {Carapella}, \citenamefont {Filatrella}, \citenamefont {Giachero},\ and\ \citenamefont {Pagano}}]{Pagano_SHG}%
  \BibitemOpen
  \bibfield  {author} {\bibinfo {author} {\bibfnamefont {C.}~\bibnamefont {Guarcello}}, \bibinfo {author} {\bibfnamefont {C.}~\bibnamefont {Barone}}, \bibinfo {author} {\bibfnamefont {G.}~\bibnamefont {Carapella}}, \bibinfo {author} {\bibfnamefont {G.}~\bibnamefont {Filatrella}}, \bibinfo {author} {\bibfnamefont {A.}~\bibnamefont {Giachero}}, \ and\ \bibinfo {author} {\bibfnamefont {S.}~\bibnamefont {Pagano}},\ }\href {\doibase 10.1063/5.0262555} {\bibfield  {journal} {\bibinfo  {journal} {Applied Physics Letters}\ }\textbf {\bibinfo {volume} {126}},\ \bibinfo {pages} {162602} (\bibinfo {year} {2025})}\BibitemShut {NoStop}%
\bibitem [{\citenamefont {Guarcello}\ \emph {et~al.}(2024)\citenamefont {Guarcello}, \citenamefont {Barone}, \citenamefont {Carapella}, \citenamefont {Granata}, \citenamefont {Filatrella}, \citenamefont {Giachero},\ and\ \citenamefont {Pagano}}]{Pagano_TWPA2}%
  \BibitemOpen
  \bibfield  {author} {\bibinfo {author} {\bibfnamefont {C.}~\bibnamefont {Guarcello}}, \bibinfo {author} {\bibfnamefont {C.}~\bibnamefont {Barone}}, \bibinfo {author} {\bibfnamefont {G.}~\bibnamefont {Carapella}}, \bibinfo {author} {\bibfnamefont {V.}~\bibnamefont {Granata}}, \bibinfo {author} {\bibfnamefont {G.}~\bibnamefont {Filatrella}}, \bibinfo {author} {\bibfnamefont {A.}~\bibnamefont {Giachero}}, \ and\ \bibinfo {author} {\bibfnamefont {S.}~\bibnamefont {Pagano}},\ }\href {\doibase https://doi.org/10.1016/j.chaos.2024.115598} {\bibfield  {journal} {\bibinfo  {journal} {Chaos, Solitons I\& Fractals}\ }\textbf {\bibinfo {volume} {189}},\ \bibinfo {pages} {115598} (\bibinfo {year} {2024})}\BibitemShut {NoStop}%
\bibitem [{\citenamefont {Malnou}\ \emph {et~al.}(2024)\citenamefont {Malnou}, \citenamefont {Miller}, \citenamefont {Estrada}, \citenamefont {Genter}, \citenamefont {Cicak}, \citenamefont {Teufel}, \citenamefont {Aumentado},\ and\ \citenamefont {Lecocq}}]{malnou2024RPMTWPA}%
  \BibitemOpen
  \bibfield  {author} {\bibinfo {author} {\bibfnamefont {M.}~\bibnamefont {Malnou}}, \bibinfo {author} {\bibfnamefont {B.~T.}\ \bibnamefont {Miller}}, \bibinfo {author} {\bibfnamefont {J.~A.}\ \bibnamefont {Estrada}}, \bibinfo {author} {\bibfnamefont {K.}~\bibnamefont {Genter}}, \bibinfo {author} {\bibfnamefont {K.}~\bibnamefont {Cicak}}, \bibinfo {author} {\bibfnamefont {J.~D.}\ \bibnamefont {Teufel}}, \bibinfo {author} {\bibfnamefont {J.}~\bibnamefont {Aumentado}}, \ and\ \bibinfo {author} {\bibfnamefont {F.}~\bibnamefont {Lecocq}},\ }\href {https://arxiv.org/abs/2406.19476} {\enquote {\bibinfo {title} {A traveling-wave parametric amplifier and converter},}\ } (\bibinfo {year} {2024}),\ \Eprint {http://arxiv.org/abs/2406.19476} {arXiv:2406.19476 [quant-ph]} \BibitemShut {NoStop}%
\bibitem [{\citenamefont {Ranadive}\ \emph {et~al.}(2024)\citenamefont {Ranadive}, \citenamefont {Fazliji}, \citenamefont {Gal}, \citenamefont {Cappelli}, \citenamefont {Butseraen}, \citenamefont {Bonet}, \citenamefont {Eyraud}, \citenamefont {Böhling}, \citenamefont {Planat}, \citenamefont {Metelmann},\ and\ \citenamefont {Roch}}]{TWPAI}%
  \BibitemOpen
  \bibfield  {author} {\bibinfo {author} {\bibfnamefont {A.}~\bibnamefont {Ranadive}}, \bibinfo {author} {\bibfnamefont {B.}~\bibnamefont {Fazliji}}, \bibinfo {author} {\bibfnamefont {G.~L.}\ \bibnamefont {Gal}}, \bibinfo {author} {\bibfnamefont {G.}~\bibnamefont {Cappelli}}, \bibinfo {author} {\bibfnamefont {G.}~\bibnamefont {Butseraen}}, \bibinfo {author} {\bibfnamefont {E.}~\bibnamefont {Bonet}}, \bibinfo {author} {\bibfnamefont {E.}~\bibnamefont {Eyraud}}, \bibinfo {author} {\bibfnamefont {S.}~\bibnamefont {Böhling}}, \bibinfo {author} {\bibfnamefont {L.}~\bibnamefont {Planat}}, \bibinfo {author} {\bibfnamefont {A.}~\bibnamefont {Metelmann}}, \ and\ \bibinfo {author} {\bibfnamefont {N.}~\bibnamefont {Roch}},\ }\href {https://arxiv.org/abs/2406.19752} {\enquote {\bibinfo {title} {A traveling wave parametric amplifier isolator},}\ } (\bibinfo {year} {2024}),\ \Eprint {http://arxiv.org/abs/2406.19752} {arXiv:2406.19752 [quant-ph]} \BibitemShut {NoStop}%
\bibitem [{\citenamefont {Delport}(2018)}]{JoSIMSite}%
  \BibitemOpen
  \bibfield  {author} {\bibinfo {author} {\bibfnamefont {J.~A.}\ \bibnamefont {Delport}},\ }\href {https://joeydelp.github.io/JoSIM/} {\enquote {\bibinfo {title} {Josim - superconducting circuit simulator},}\ } (\bibinfo {year} {2018})\BibitemShut {NoStop}%
\bibitem [{\citenamefont {Delport}\ \emph {et~al.}(2019)\citenamefont {Delport}, \citenamefont {Jackman}, \citenamefont {le~Roux},\ and\ \citenamefont {Fourie}}]{Delport2019}%
  \BibitemOpen
  \bibfield  {author} {\bibinfo {author} {\bibfnamefont {J.}~\bibnamefont {Delport}}, \bibinfo {author} {\bibfnamefont {K.}~\bibnamefont {Jackman}}, \bibinfo {author} {\bibfnamefont {P.}~\bibnamefont {le~Roux}}, \ and\ \bibinfo {author} {\bibfnamefont {C.}~\bibnamefont {Fourie}},\ }\href {\doibase 10.1109/TASC.2019.2897312} {\bibfield  {journal} {\bibinfo  {journal} {IEEE Transactions on Applied Superconductivity}\ }\textbf {\bibinfo {volume} {PP}},\ \bibinfo {pages} {1} (\bibinfo {year} {2019})}\BibitemShut {NoStop}%
\bibitem [{Spi()}]{Spice_cite}%
  \BibitemOpen
  \href {http://www.wrcad.com/wrspice.html} {\enquote {\bibinfo {title} {Whiteley research inc., wrspice circuit simulator},}\ }\BibitemShut {NoStop}%
\bibitem [{\citenamefont {Whiteley}(1991)}]{133816}%
  \BibitemOpen
  \bibfield  {author} {\bibinfo {author} {\bibfnamefont {S.}~\bibnamefont {Whiteley}},\ }\href {\doibase 10.1109/20.133816} {\bibfield  {journal} {\bibinfo  {journal} {IEEE Transactions on Magnetics}\ }\textbf {\bibinfo {volume} {27}},\ \bibinfo {pages} {2902} (\bibinfo {year} {1991})}\BibitemShut {NoStop}%
\bibitem [{\citenamefont {Hutter}\ \emph {et~al.}(2011)\citenamefont {Hutter}, \citenamefont {Thol\'en}, \citenamefont {Stannigel}, \citenamefont {Lidmar},\ and\ \citenamefont {Haviland}}]{Tholen}%
  \BibitemOpen
  \bibfield  {author} {\bibinfo {author} {\bibfnamefont {C.}~\bibnamefont {Hutter}}, \bibinfo {author} {\bibfnamefont {E.~A.}\ \bibnamefont {Thol\'en}}, \bibinfo {author} {\bibfnamefont {K.}~\bibnamefont {Stannigel}}, \bibinfo {author} {\bibfnamefont {J.}~\bibnamefont {Lidmar}}, \ and\ \bibinfo {author} {\bibfnamefont {D.~B.}\ \bibnamefont {Haviland}},\ }\href {\doibase 10.1103/PhysRevB.83.014511} {\bibfield  {journal} {\bibinfo  {journal} {Phys. Rev. B}\ }\textbf {\bibinfo {volume} {83}},\ \bibinfo {pages} {014511} (\bibinfo {year} {2011})}\BibitemShut {NoStop}%
\bibitem [{\citenamefont {Malnou}\ \emph {et~al.}(2021)\citenamefont {Malnou}, \citenamefont {Vissers}, \citenamefont {Wheeler}, \citenamefont {Aumentado}, \citenamefont {Hubmayr}, \citenamefont {Ullom},\ and\ \citenamefont {Gao}}]{Malnou_TWPA_CME}%
  \BibitemOpen
  \bibfield  {author} {\bibinfo {author} {\bibfnamefont {M.}~\bibnamefont {Malnou}}, \bibinfo {author} {\bibfnamefont {M.}~\bibnamefont {Vissers}}, \bibinfo {author} {\bibfnamefont {J.}~\bibnamefont {Wheeler}}, \bibinfo {author} {\bibfnamefont {J.}~\bibnamefont {Aumentado}}, \bibinfo {author} {\bibfnamefont {J.}~\bibnamefont {Hubmayr}}, \bibinfo {author} {\bibfnamefont {J.}~\bibnamefont {Ullom}}, \ and\ \bibinfo {author} {\bibfnamefont {J.}~\bibnamefont {Gao}},\ }\href {\doibase 10.1103/PRXQuantum.2.010302} {\bibfield  {journal} {\bibinfo  {journal} {PRX Quantum}\ }\textbf {\bibinfo {volume} {2}},\ \bibinfo {pages} {010302} (\bibinfo {year} {2021})}\BibitemShut {NoStop}%
\bibitem [{\citenamefont {Chaudhuri}, \citenamefont {Gao},\ and\ \citenamefont {Irwin}(2014)}]{Side_bands}%
  \BibitemOpen
  \bibfield  {author} {\bibinfo {author} {\bibfnamefont {S.}~\bibnamefont {Chaudhuri}}, \bibinfo {author} {\bibfnamefont {J.}~\bibnamefont {Gao}}, \ and\ \bibinfo {author} {\bibfnamefont {K.}~\bibnamefont {Irwin}},\ }\href@noop {} {\bibfield  {journal} {\bibinfo  {journal} {IEEE Transactions on Applied Superconductivity}\ }\textbf {\bibinfo {volume} {25}} (\bibinfo {year} {2014})}\BibitemShut {NoStop}%
\bibitem [{\citenamefont {Pozar}(2005)}]{Pozar:882338}%
  \BibitemOpen
  \bibfield  {author} {\bibinfo {author} {\bibfnamefont {D.~M.}\ \bibnamefont {Pozar}},\ }\href {https://cds.cern.ch/record/882338} {\emph {\bibinfo {title} {{Microwave engineering; 3rd ed.}}}}\ (\bibinfo  {publisher} {Wiley},\ \bibinfo {address} {Hoboken, NJ},\ \bibinfo {year} {2005})\BibitemShut {NoStop}%
\bibitem [{\citenamefont {Gal}\ \emph {et~al.}(2025)\citenamefont {Gal}, \citenamefont {Butseraen}, \citenamefont {Ranadive}, \citenamefont {Cappelli}, \citenamefont {Fazliji}, \citenamefont {Bonet}, \citenamefont {Eyraud}, \citenamefont {Planat},\ and\ \citenamefont {Roch}}]{gal2025gaincompressionTWPA}%
  \BibitemOpen
  \bibfield  {author} {\bibinfo {author} {\bibfnamefont {G.~L.}\ \bibnamefont {Gal}}, \bibinfo {author} {\bibfnamefont {G.}~\bibnamefont {Butseraen}}, \bibinfo {author} {\bibfnamefont {A.}~\bibnamefont {Ranadive}}, \bibinfo {author} {\bibfnamefont {G.}~\bibnamefont {Cappelli}}, \bibinfo {author} {\bibfnamefont {B.}~\bibnamefont {Fazliji}}, \bibinfo {author} {\bibfnamefont {E.}~\bibnamefont {Bonet}}, \bibinfo {author} {\bibfnamefont {E.}~\bibnamefont {Eyraud}}, \bibinfo {author} {\bibfnamefont {L.}~\bibnamefont {Planat}}, \ and\ \bibinfo {author} {\bibfnamefont {N.}~\bibnamefont {Roch}},\ }\href {https://arxiv.org/abs/2502.03022} {\enquote {\bibinfo {title} {Gain compression in josephson traveling-wave parametric amplifiers},}\ } (\bibinfo {year} {2025}),\ \Eprint {http://arxiv.org/abs/2502.03022} {arXiv:2502.03022 [quant-ph]} \BibitemShut {NoStop}%
\bibitem [{Sol()}]{Solve_ivp_cite}%
  \BibitemOpen
  \href {https://docs.scipy.org/doc/scipy/reference/generated/scipy.integrate.solve_ivp.html} {\enquote {\bibinfo {title} {Scipy documentation},}\ }\BibitemShut {NoStop}%
\bibitem [{\citenamefont {Bogacki}\ and\ \citenamefont {Shampine}(1989)}]{RK23}%
  \BibitemOpen
  \bibfield  {author} {\bibinfo {author} {\bibfnamefont {P.}~\bibnamefont {Bogacki}}\ and\ \bibinfo {author} {\bibfnamefont {L.}~\bibnamefont {Shampine}},\ }\href {\doibase https://doi.org/10.1016/0893-9659(89)90079-7} {\bibfield  {journal} {\bibinfo  {journal} {Applied Mathematics Letters}\ }\textbf {\bibinfo {volume} {2}},\ \bibinfo {pages} {321} (\bibinfo {year} {1989})}\BibitemShut {NoStop}%
\bibitem [{\citenamefont {Dormand}\ and\ \citenamefont {Prince}(1980)}]{RK45}%
  \BibitemOpen
  \bibfield  {author} {\bibinfo {author} {\bibfnamefont {J.}~\bibnamefont {Dormand}}\ and\ \bibinfo {author} {\bibfnamefont {P.}~\bibnamefont {Prince}},\ }\href {\doibase https://doi.org/10.1016/0771-050X(80)90013-3} {\bibfield  {journal} {\bibinfo  {journal} {Journal of Computational and Applied Mathematics}\ }\textbf {\bibinfo {volume} {6}},\ \bibinfo {pages} {19} (\bibinfo {year} {1980})}\BibitemShut {NoStop}%
\bibitem [{\citenamefont {Hairer}, \citenamefont {Norsett},\ and\ \citenamefont {Wanner}(1993)}]{RK8}%
  \BibitemOpen
  \bibfield  {author} {\bibinfo {author} {\bibfnamefont {E.}~\bibnamefont {Hairer}}, \bibinfo {author} {\bibfnamefont {S.}~\bibnamefont {Norsett}}, \ and\ \bibinfo {author} {\bibfnamefont {G.}~\bibnamefont {Wanner}},\ }\href {\doibase 10.1007/978-3-540-78862-1} {\emph {\bibinfo {title} {Solving Ordinary Differential Equations I: Nonstiff Problems}}},\ Vol.~\bibinfo {volume} {8}\ (\bibinfo {year} {1993})\BibitemShut {NoStop}%
\bibitem [{\citenamefont {Shampine}\ and\ \citenamefont {Reichelt}(1997)}]{BDF}%
  \BibitemOpen
  \bibfield  {author} {\bibinfo {author} {\bibfnamefont {L.~F.}\ \bibnamefont {Shampine}}\ and\ \bibinfo {author} {\bibfnamefont {M.~W.}\ \bibnamefont {Reichelt}},\ }\href {\doibase 10.1137/S1064827594276424} {\bibfield  {journal} {\bibinfo  {journal} {SIAM Journal on Scientific Computing}\ }\textbf {\bibinfo {volume} {18}},\ \bibinfo {pages} {1} (\bibinfo {year} {1997})}\BibitemShut {NoStop}%
\bibitem [{\citenamefont {Rizvanov}(2025)}]{CMT_github}%
  \BibitemOpen
  \bibfield  {author} {\bibinfo {author} {\bibfnamefont {E.}~\bibnamefont {Rizvanov}},\ }\href@noop {} {\enquote {\bibinfo {title} {Coupled mode theory simulator for ptwpa},}\ }\bibinfo {howpublished} {\url{https://github.com/EmilRizvanov/CMT3}} (\bibinfo {year} {2025}),\ \bibinfo {note} {gitHub repository}\BibitemShut {NoStop}%
\end{thebibliography}%


\end{document}